\documentclass{article}
\usepackage{slashed,verbatim,subfigure}
\usepackage[numbers,sort&compress]{natbib}
\usepackage{amsmath}
\usepackage{amssymb}
\usepackage{appendix}
\usepackage{graphicx}
\usepackage{dcolumn}
\usepackage{bm}
\usepackage[colorlinks=true,linktocpage=true,
linkcolor=blue,citecolor=blue]{hyperref}
\usepackage[a4paper]{geometry}
\newcommand{\nd}{\noindent}
\newcommand{\be}{\begin{eqnarray}}
\newcommand{\ee}{\end{eqnarray}}

\begin{document}
\large
\title{\bf{Momentum transport properties of a hot and dense QCD matter in a weak magnetic field}}
\author{Shubhalaxmi Rath\footnote{shubhalaxmi@iitb.ac.in}~~and~~Sadhana Dash\footnote{sadhana@phy.iitb.ac.in}\vspace{0.03in} \\ Department of Physics, Indian Institute of Technology Bombay, Mumbai 400076, India}
\date{}
\maketitle
\begin{abstract}
We have studied the momentum transport properties of a hot and dense QCD matter 
in the presence of weak magnetic field by determining the shear ($\eta$) and 
bulk ($\zeta$) viscosities in the relaxation time approximation of kinetic theory. The dependence of $\eta$ and $\zeta$ on the temperature has been explored in the presence of weak magnetic field ($B$-field) and finite chemical potential ($\mu$). It is observed that both shear and bulk viscosities get decreased in the presence of a weak magnetic field, whereas the finite chemical potential increases these viscosities, specifically at low temperatures. This study is important to understand the sound attenuation through the Prandtl number (Pr), the nature of the flow through the Reynolds number (Re), the fluidity and location of transition 
point of the matter through the ratios $\eta/s$ and $\zeta/s$ ($s$ is the entropy density), respectively. The Prandtl number is observed  to increase in the weak magnetic field, whereas the  presence of 
a finite chemical potential reduces its magnitude as compared to the scenario of absence of $B$-field and $\mu$. However, Pr still remains 
larger than unity, indicating that the energy dissipation due to the sound attenuation is mostly governed by the momentum diffusion. It is noticed that the weak magnetic field 
makes the Reynolds number larger, whereas the chemical potential makes it smaller 
than that in the absence of $B$-field and $\mu$. We have observed that the 
ratio $\eta/s$  decreases in the weak magnetic field regime, whereas the 
finite chemical potential increases its value, but the ratio $\zeta/s$ is found to 
decrease in the presence of weak magnetic field as well as finite chemical potential. 

\end{abstract}

\newpage

\section{Introduction}
The ultrarelativistic heavy ion collisions (URHICs) at Relativistic Heavy Ion 
Collider (RHIC) and Large Hadron Collider (LHC) have provided strong evidence
of the formation of a strongly interacting matter, known as the quark-gluon plasma 
(QGP). One of the salient goals of these experiments is to divulge the 
transport properties of the QGP. The transport coefficients are sensitive to the 
relevant degrees of freedom and their respective interactions within the QGP medium. 
For example, the shear viscosity ($\eta$) gives information about the momentum transfer 
in the presence of inhomogeneity of fluid velocity and the bulk viscosity ($\zeta$) delineates the change of local pressure due to either contraction or expansion of fluid. Shear viscosity is related to a change in shape at constant volume, whereas bulk viscosity is related to a change in volume at constant shape. The dimensionless ratios of shear and bulk viscosities ($\eta/s$ and $\zeta/s$) to entropy density ($s$) characterize the intrinsic ability of a system to relax towards  equilibrium. The exploration of shear and bulk viscosities reveal about the fluid dynamical behavior of the medium. The shear and bulk viscosity calculations were performed through various approaches, such as the relativistic Boltzmann transport equation in the relaxation time approximation \cite{Danielewicz:PRD31'1985,Heckmann:EPJA48'2012,Yasui:PRD96'2017}, the Green-Kubo formula\cite{Basagoiti:PRD66'2002,Kharzeev:JHEP0809'2008,Moore:JHEP0809'2008,
Plumari:PRC86'2012}, the lattice simulations \cite{Astrakhantsev:JHEP1704'2017,Astrakhantsev:PRD98'2018}, the 
molecular dynamics simulation \cite{Gelman:PRC74'2006}, the perturbation theory \cite{Arnold:JHEP'2000'2003,Arnold:PRD74'2006,Hidaka:PRD78'2008} etc. The $N=4$ supersymmetric $SU(N_c)$ Yang-Mills theory has estimated the lower bound of the ratio $\eta/s$ as close to $1/(4\pi)$, which is also known as the Kovtun-Son-Starinets (KSS) lower bound \cite{Kovtun:PRL94'2005}. This estimated lower bound has been conjectured to be the lower bound of $\eta/s$ for different physical systems, such as helium, nitrogen and water at pressures 0.1 MPa, 10 MPa, and 100 MPa, respectively. The transition from hadrons to quark-gluon plasma has a similar behavior in the ratio $\eta/s$ \cite{Csernai:PRL97'2006}. 

In addition, the Au$-$Au collision at RHIC has also reported very low value of $\eta/s\sim1/(4\pi)$ for the QGP medium formed indicating that the hot and dense matter produced in URHICs behaves
like a perfect fluid. Using lattice gauge theory principles, $\eta/s$ has been studied for  few values of temperature in pure Yang-Mills theory \cite{Nakamura:PRL94'2005,Meyer:PRD76'2007,Astrakhantsev:JHEP1704'2017}. According to the  Yang-Mills theory and perturbative QCD \cite{Arnold:JHEP0305'2003}, an increase in $\eta/s$ is observed in the presence of dynamical quarks \cite{Astrakhantsev:JHEP1704'2017,Meyer:NPA830'2009}. Through the functional diagrammatic approach to QCD, references \cite{Haas:PRD90'2014,Christiansen:PRL115'2015} have evaluated $\eta$ in Yang-Mills theory and the results are in good agreement with the lattice results for QCD with (2+1)-quark flavors. Both the  approaches have observed a minimal $\eta/s$ of about 0.2 near the phase transition temperature, which is slightly larger than $1/(4\pi)$. Similar results were also obtained in perturbation theory \cite{Christiansen:PRL115'2015,Jackson:JPG45'2018}. For massless QGP, the bulk viscosity is very small compared to the shear viscosity for which it was neglected by some early viscous 
hydrodynamic simulations in the dissipative part of the energy-momentum tensor \cite{Muronga:PRL88'2002,Song:JPG36'2009}. The vanishing of the ratio $\zeta/s$ explains the restoration of chiral symmetry of the matter. On the other hand, a sharp rise of the ratio $\zeta/s$ in the vicinity of the phase transition temperature of matter is reported in ref. \cite{Kharzeev:JHEP0809'2008}. Large value of the bulk viscosity signifies large fluctuations in the pressure. Although $\zeta$ vanishes for QGP with massless flavors at the classical level, but the quantum effects break the conformal symmetry of QCD, thus generating a nonzero bulk viscosity, which is described by the lattice calculation in the $SU(3)$ gauge theory \cite{Meyer:PRL100'2008}. As a result, the ratio $\zeta/s$ acts as a measure of the deviation of the strongly interacting matter from conformality. 

The aforementioned estimations were made for most central collisions. However, in noncentral events of heavy ion collisions, when two nuclei travelling with 
ultrarelativistic speeds collide with each other, an intensely strong magnetic field perpendicular to the collision plane is expected to be produced at very early stages. Depending on the centrality, the strength of the magnetic field may vary between $m_\pi^2$ (${10}^{18}$ Gauss) at RHIC and 15 $m_\pi^2$ at LHC \cite{Skokov:IJMPA24'2009} and at extreme cases it may reach 50 $m_\pi^2$. The magnetic field is very strong for very short duration and becomes weak. So, there are two limits: strong magnetic field and weak magnetic field. In the strong magnetic field limit, the energy scale associated with the magnetic field is greater than the energy scale related to the temperature ($|q_fB| \gg T^2$, where $|q_f|$ is the absolute electronic charge of quark with flavor $f$). On the other hand, in the weak magnetic field limit, the energy scale associated with the magnetic field is smaller than the energy scale related to the temperature ($|q_fB| \ll T^2$). According to some observations \cite{Tuchin:PRC82'2010,Rath:PRD100'2019}, the lifetime of such magnetic field gets significantly extended in an electrically conducting medium and is comparable with the lifetime of the partonic medium. In addition, high baryon densities are expected to be evidenced in Compressed Baryonic Matter (CBM) experiment at Facility for Antiproton and Ion Research (FAIR) and Nuclotron-based Ion Collider fAcility (NICA) at Joint Institute for Nuclear Research (JINR) in fixed target experiments. Thus, the shear viscosity, the bulk viscosity and the associated transport properties of the medium are prone to be altered by the presence of both magnetic field and chemical potential. Previously, the effects of magnetic field on the QCD thermodynamics \cite{Rath:JHEP1712'2017,Bandyopadhyay:PRD100'2019,Rath:EPJA55'2019,Karmakar:PRD99'2019}, the heavy quark diffusion \cite{Fukushima:PRD93'2016}, the conductive properties \cite{Hattori:PRD94'2016,Feng:PRD96'2017,Kurian:PRD96'2017,Fukushima:PRL120'2018,
Rath:PRD100'2019,Rath:EPJC80'2020,Thakur:PRD100'2019,Kurian:EPJC79'2019}, the magnetohydrodynamics \cite{Roy:PLB750'2015,Inghirami:EPJC76'2016}, the photon and dilepton productions from QGP \cite{Hees:PRC84'2011,Tuchin:PRC88'2013,Mamo:JHEP1308'2013,Shen:PRC89'2014}, etc. have been explored. Recently, the collective effects of weak magnetic field and finite chemical potential on the charge transport, the heat transport and some related transport coefficients were explored in ref. \cite{Rath:2112.11802}. Viscous properties were also studied previously by using different models and approximations at finite magnetic field. For example, in ref. \cite{Nam:PRD87'2013} authors had employed the diluted instanton liquid model and the Green-Kubo formula to study the shear viscosity of the SU(2) light-flavor quark matter at finite temperature under the strong magnetic field limit. In ref. \cite{Hattori:PRD96'2017} authors had investigated the viscosities of the quark-gluon plasma in the presence of the strong magnetic field with the leading-log and lowest Landau level (LLL) approximations. Authors in ref. \cite{Li:PRD97'2018} had computed the shear viscosity of two-flavor QCD plasma in a magnetic field by using the perturbative QCD at leading log order. In ref. \cite{Denicol:PRD98'2018} authors had investigated the viscosities using the nonresistive dissipative magnetohydrodynamics from the Boltzmann equation in the 14-moment approximation at finite magnetic field. Authors in ref. \cite{Das:PRD100'2019} had estimated viscosities using the relativistic Boltzmann transport equation in the relaxation time approximation, but for a hot and dense hadronic matter. In ref. \cite{Rath:PRD102'2020} authors had investigated the effects of the strong magnetic field-induced and asymptotic expansion-induced anisotropies on viscosities for a hot QCD matter using the kinetic theory approach, while in ref. \cite{Rath:EPJC81'2021}, the effects of the strong magnetic field and density on viscosities had been explored. In the present work, (i) we have studied shear and bulk viscosities for a hot QCD matter in the presence of both magnetic field and finite chemical potential. We have estimated the viscosities by solving the relativistic Boltzmann transport equation in the kinetic theory approach and used the weak magnetic field limit, where the energy scale associated with the temperature is larger than the energy scale related to the magnetic field, {\em i.e.} $T^2\gg |q_fB|$. So, we have used the ansatz method in the weak magnetic field limit to calculate viscosities in the first part of section 2, where the terms containing $\omega_c$ (cyclotron frequency) and its higher orders have been neglected. (ii) In the second part of section 2, we have revisited the viscosity coefficients in the general configuration of magnetic field (no weak or strong magnetic field limit) and observed how they are related to the viscosities calculated using the ansatz method. (iii) We have extended our study to know the collective effects of weak magnetic field and density on some applications of viscosities, such as the Prandtl number (Pr), the Reynolds number (Re), specific shear viscosity ($\eta/s$) and specific bulk viscosity ($\zeta/s$). (iv) We have used the quasiparticle model, wherein the interactions among the medium constituents have been incorporated through the thermal masses of particles. 

The present work is organized as follows. In Section 2, the momentum transport properties have been studied by deriving the response functions, {\em viz.} the shear viscosity and the bulk viscosity in the kinetic theory approach with a short description of the quasiparticle model. The results are presented in Section 3 while Section 4 discusses some applications of both the viscosities in terms of the Prandtl number, the Reynolds number and the ratios $\eta/s$ and $\zeta/s$. The work is summarized in Section 5. 

\section{Momentum transport properties}
A fluid system slightly shifted from its equilibrium state due to the nonuniformity of its constituent flow with respect to the macroscopic velocity, can possess finite shear and bulk viscosities. We calculate the viscosities by assuming a local temperature $T(x)$ and flow velocity $u^\mu(x)$. For a nonequilibrium system, the dissipative part of the energy-momentum tensor $\Delta T^{\mu\nu}$ is written in terms of the equilibrium energy-momentum tensor $T_{(0)}^{\mu\nu}$ as
\begin{eqnarray}
\Delta T^{\mu\nu}=T^{\mu\nu}-T_{(0)}^{\mu\nu}
~.\end{eqnarray}
For the partonic system, $\Delta T^{\mu\nu}$ can also be written in terms 
of the infinitesimal changes of the quark, antiquark and gluon 
distribution functions as
\be\label{em1}
\Delta T^{\mu\nu}=\int\frac{d^3{\rm p}}{(2\pi)^3}p^\mu p^\nu \left[\sum_f g_f\frac{\left(\delta f_f+\delta \bar{f}_f\right)}{{\omega_f}}+g_g\frac{\delta f_g}{\omega_g}\right]
,\ee
where `$f$' represents the flavor index for three flavors $u$, $d$ and $s$. In eq. \eqref{em1}, $g_f$ and $\delta f_f$ ($\delta \bar{f_f}$) denote the degeneracy factor and the infinitesimal change in the quark (antiquark) distribution function of $f$th flavor, respectively. For the gluon, $g_g$ and $\delta f_g$ denote the degeneracy factor and the infinitesimal change in its distribution function, respectively. The infinitesimal changes in quark, antiquark and gluon distribution functions are defined as $\delta f_f=f_f-f_f^0$, $\delta \bar{f}_f=\bar{f}_f-\bar{f}_f^0$ and $\delta f_g=f_g-f_g^0$, respectively. Here, $f_f^0$, $\bar{f}_f^0$ and $f_g^0$ are the equilibrium distribution functions for quark, antiquark and gluon, respectively, which have the following forms, 
\be\label{Q.D.F.}
&&f_f^0=\frac{1}{e^{\beta \left(u^\alpha p_\alpha-\mu_f\right)}+1} ~, \\
&&\label{Q.D.F.1}\bar{f}_f^0=\frac{1}{e^{\beta \left(u^\alpha p_\alpha+\mu_f\right)}+1} ~, \\
&&\label{G.D.F.}f_g^0=\frac{1}{e^{\beta u^\alpha p_\alpha}-1}
~,\ee
where $T=\beta^{-1}$, $u^\alpha$ denotes the four-velocity 
of fluid and $\mu_f$ represents the chemical potential of $f$th flavor of quark. In above equations, for quark and antiquark, $p_\alpha\equiv\left(\omega_f,\mathbf{p}\right)$ with 
$\omega_f=\sqrt{\mathbf{p}^2+m_f^2}$ and for gluon, $p_\alpha\equiv\left(\omega_g,\mathbf{p}\right)$. In order to determine the infinitesimal change in the particle distribution function, we are going to solve the relativistic Boltzmann transport equation in the relaxation time approximation for finite magnetic field and chemical potential, 
\be\label{R.B.T.E.}
p^\mu\frac{\partial f_f(x,p)}{\partial x^\mu}+\mathcal{F}^\mu\frac{\partial f_f(x,p)}{\partial p^\mu}=-\frac{p_\nu u^\nu}{\tau_f}\delta f_f(x,p)
~,\ee
where $f_f=\delta f_f+f_f^0$. The external force $\mathcal{F}^\mu=qF^{\mu\nu}p_\nu=(p^0\mathbf{v}\cdot\mathbf{F}, p^0\mathbf{F})$, where $F^{\mu\nu}$ represents the electromagnetic field strength tensor and $\mathbf{F}$ denotes the Lorentz force, $\mathbf{F}=q(\mathbf{E}+\mathbf{v}\times\mathbf{B})$. The relations between the components of $F^{\mu\nu}$ and the components of electric and 
magnetic fields are given by $F^{0i}=E^i$, $F^{i0}=-E^i$ 
and $F^{ij}=\frac{1}{2}\epsilon^{ijk}B_k$. The relaxation times for quarks (antiquarks), $\tau_f$ ($\tau_{\bar{f}}$) and for gluons, $\tau_g$ are respectively written \cite{Hosoya:NPB250'1985} as
\begin{eqnarray}
&&\tau_{f(\bar{f})}=\frac{1}{5.1T\alpha_s^2\log\left(1/\alpha_s\right)\left[1+0.12 
(2N_f+1)\right]} ~, \label{Q.R.T.} \\
&&\tau_g=\frac{1}{22.5T\alpha_s^2\log\left(1/\alpha_s\right)\left[1+0.06N_f
\right]} \label{G.R.T.}
~.\end{eqnarray}
To solve eq. \eqref{R.B.T.E.}, we take the following ansatz which was first suggested by ref. \cite{Feng:PRD96'2017}, 
\be\label{ansatz}
f_f=f_f^0-\tau_fq\mathbf{E}\cdot\frac{\partial f_f^0}{\partial \mathbf{p}}-\mathbf{\Gamma}\cdot\frac{\partial f_f^0}{\partial \mathbf{p}}
~,\ee
where $\mathbf{\Gamma}$ is associated with the magnetic 
field. The partial derivatives in the above ansatz are 
evaluated as
\begin{eqnarray*}
&&\frac{\partial f_f^0}{\partial p_x}=-\beta v_xf_f^0\left(1-f_f^0\right), ~~ \frac{\partial f_f^0}{\partial p_y}=-\beta v_yf_f^0\left(1-f_f^0\right), ~~ \frac{\partial f_f^0}{\partial p_z}=-\beta v_zf_f^0\left(1-f_f^0\right)
~.\end{eqnarray*}
Assuming that the electric field is along x-direction ($\mathbf{E}=E\hat{x}$) and the magnetic field is along z-direction ($\mathbf{B}=B\hat{z}$), the relativistic 
Boltzmann transport equation \eqref{R.B.T.E.} using the ansatz \eqref{ansatz} can be rewritten as
\be\label{eq.1}
\frac{\tau_f}{p_0}p^\mu\frac{\partial f_f^0}{\partial x^\mu}+\beta f_f^0\left(1-f_f^0\right)\left(\Gamma_xv_x+\Gamma_yv_y+\Gamma_zv_z\right)+\tau_fqEv_x\frac{\partial f_f}{\partial p_0}-qB\tau_f\left(v_x\frac{\partial f_f}{\partial p_y}-v_y\frac{\partial f_f}{\partial p_x}\right)=0
.\ee
The partial derivatives in the above equation are calculated as
\be
\nonumber v_x\frac{\partial f_f}{\partial p_0} &=& -\beta v_xf_f^0\left(1-f_f^0\right)-qE\tau_f\beta f_f^0\left(1-f_f^0\right)v_x^2\left(\frac{1}{\omega_f}+\beta\right) \\ && \nonumber -\beta f_f^0\left(1-f_f^0\right)\Gamma_xv_x^2\left(\frac{1}{\omega_f}+\beta\right) \\ && -\beta f_f^0\left(1-f_f^0\right)\Gamma_yv_xv_y\left(\frac{1}{\omega_f}+\beta\right)-\beta f_f^0\left(1-f_f^0\right)\Gamma_zv_xv_z\left(\frac{1}{\omega_f}+\beta\right), \\
\nonumber v_x\frac{\partial f_f}{\partial p_y} &=& -\beta v_xv_yf_f^0\left(1-f_f^0\right)-qE\tau_f\beta f_f^0\left(1-f_f^0\right)v_x^2v_y\left(\frac{1}{\omega_f}+\beta\right) \\ && \nonumber -\beta f_f^0\left(1-f_f^0\right)\Gamma_xv_x^2v_y\left(\frac{1}{\omega_f}+\beta\right)-\beta f_f^0\left(1-f_f^0\right)\Gamma_yv_xv_y^2\left(\frac{1}{\omega_f}+\beta\right) \\ && +\frac{v_x\Gamma_y\beta f_f^0\left(1-f_f^0\right)}{\omega_f}-\beta f_f^0\left(1-f_f^0\right)\Gamma_zv_xv_yv_z\left(\frac{1}{\omega_f}+\beta\right) , \\
\nonumber v_y\frac{\partial f_f}{\partial p_x} &=& -\beta v_yv_xf_f^0\left(1-f_f^0\right)-qE\tau_f\beta f_f^0\left(1-f_f^0\right)v_yv_x^2\left(\frac{1}{\omega_f}+\beta\right)+\frac{qE\tau_f\beta f_f^0\left(1-f_f^0\right)v_y}{\omega_f} \\ && \nonumber -\beta f_f^0\left(1-f_f^0\right)\Gamma_xv_yv_x^2\left(\frac{1}{\omega_f}+\beta\right)+\frac{\Gamma_x\beta f_f^0\left(1-f_f^0\right)v_y}{\omega_f} \\ && -\beta f_f^0\left(1-f_f^0\right)\Gamma_yv_y^2v_x\left(\frac{1}{\omega_f}+\beta\right)-\beta f_f^0\left(1-f_f^0\right)\Gamma_zv_yv_zv_x\left(\frac{1}{\omega_f}+\beta\right)
.\ee
Substituting the values of partial derivatives in eq. \eqref{eq.1} and then 
dropping higher order velocity terms, we obtain 
\be\label{eq.2}
&&\nonumber J-\beta f_f^0\left(1-f_f^0\right)\tau_fqEv_x+\beta f_f^0\left(1-f_f^0\right)\left(\Gamma_xv_x+\Gamma_yv_y+\Gamma_zv_z\right) \\ && -\frac{qB\tau_f\beta f_f^0\left(1-f_f^0\right)}{\omega_f}\left(v_x\Gamma_y-v_y\Gamma_x\right) +\frac{\tau_f^2qBqEv_y\beta f_f^0\left(1-f_f^0\right)}{\omega_f}=0
.\ee
In getting the above equation, we have also replaced $J=\frac{\tau_f}{p_0}p^\mu\frac{\partial f_f^0}{\partial x^\mu}$. For quark distribution function, we have calculated $J$ as
\be\label{J}
\nonumber J &=& -\beta\tau_f f_f^0\left(1-f_f^0\right)\left[\left\lbrace\omega_f\left(\frac{\partial P}{\partial \varepsilon}\right)-\frac{\rm p^2}{3\omega_f}\right\rbrace\partial_l u^l+p^l\left(\frac{\partial_l P}{\varepsilon+P}-\frac{\partial_l T}{T}\right)\right. \\ && \left.-\frac{Tp^l}{\omega_f}\partial_l\left(\frac{\mu_f}{T}\right)-\frac{p^kp^l}{2\omega_f}W_{kl}\right]
,\ee
where $W_{kl}=\partial_k u_l+\partial_l u_k-\frac{2}{3}\delta_{kl}\partial_j u^j$. Similarly for antiquark and gluon distribution functions, we get
\be
\nonumber \bar{J} &=& -\beta\tau_{\bar{f}} \bar{f_f}^0\left(1-\bar{f_f}^0\right)\left[\left\lbrace\omega_f\left(\frac{\partial P}{\partial \varepsilon}\right)-\frac{\rm p^2}{3\omega_f}\right\rbrace\partial_l u^l+p^l\left(\frac{\partial_l P}{\varepsilon+P}-\frac{\partial_l T}{T}\right)\right. \\ && \left.+\frac{Tp^l}{\omega_f}\partial_l\left(\frac{\mu_f}{T}\right)-\frac{p^kp^l}{2\omega_f}W_{kl}\right], \\ 
\nonumber J_g &=& -\beta\tau_g f_g^0\left(1+f_g^0\right)\left[\left\lbrace\omega_f\left(\frac{\partial P}{\partial \varepsilon}\right)-\frac{\rm p^2}{3\omega_g}\right\rbrace\partial_l u^l+p^l\left(\frac{\partial_l P}{\varepsilon+P}-\frac{\partial_l T}{T}\right)\right. \\ && \left.-\frac{p^kp^l}{2\omega_g}W_{kl}\right]
,\ee
respectively. With the help of equations \eqref{eq.2}, \eqref{J} and \eqref{ansatz}, 
we get the nonequilibrium part of the quark distribution function 
(determined in appendix \ref{I.C. of Q.D.F.}) as
\be\label{deltaf.1}
\nonumber\delta f_f &=& qE\tau_fv_x\beta f_f^0\left(1-f_f^0\right)+v_x\beta f_f^0\left(1-f_f^0\right)\left[\frac{\tau_f}{1+\omega_c^2\tau_f^2}\frac{p_0p_x}{{\rm p}^2}\left\lbrace\omega_f\left(\frac{\partial P}{\partial \varepsilon}\right)-\frac{\rm p^2}{3\omega_f}\right\rbrace\partial_l u^l\right. \\ && \left.\nonumber+\frac{\omega_c\tau_f^2}{1+\omega_c^2\tau_f^2}\frac{p_0p_y}{{\rm p}^2}\left\lbrace\omega_f\left(\frac{\partial P}{\partial \varepsilon}\right)-\frac{\rm p^2}{3\omega_f}\right\rbrace\partial_l u^l-\frac{\tau_f}{1+\omega_c^2\tau_f^2}\frac{p_kW_{kx}}{2}-\frac{\omega_c\tau_f^2}{1+\omega_c^2\tau_f^2}\frac{p_kW_{ky}}{2}\right. \\ && \left.\nonumber+\frac{\left(\tau_f-\omega_c^2\tau_f^3\right)qE}{1+\omega_c^2\tau_f^2}\right]+v_y\beta f_f^0\left(1-f_f^0\right)\left[\frac{\tau_f}{1+\omega_c^2\tau_f^2}\frac{p_0p_y}{{\rm p}^2}\left\lbrace\omega_f\left(\frac{\partial P}{\partial \varepsilon}\right)-\frac{\rm p^2}{3\omega_f}\right\rbrace\partial_l u^l\right. \\ && \left.\nonumber -\frac{\omega_c\tau_f^2}{1+\omega_c^2\tau_f^2}\frac{p_0p_x}{{\rm p}^2}\left\lbrace\omega_f\left(\frac{\partial P}{\partial \varepsilon}\right)-\frac{\rm p^2}{3\omega_f}\right\rbrace\partial_l u^l-\frac{\tau_f}{1+\omega_c^2\tau_f^2}\frac{p_kW_{ky}}{2}\right. \\ && \left.+\frac{\omega_c\tau_f^2}{1+\omega_c^2\tau_f^2}\frac{p_kW_{kx}}{2}-\frac{2\omega_c\tau_f^2qE}{1+\omega_c^2\tau_f^2}\right]
,\ee
where the cyclotron frequency, $\omega_c$ is defined as $\omega_c=\frac{qB}{\omega_f}$. 
In a first order theory, for infinitesimal deviation of the system 
from its equilibrium, the spatial component of the dissipative 
part of the energy-momentum tensor is defined \cite{Lifshitz:BOOK'1981,Hosoya:NPB250'1985,Landau:BOOK'1987} as
\be\label{definition}
\Delta T^{ij}=-\eta W^{ij}-\zeta\delta^{ij}\partial_l u^l
.\ee
Here the shear viscosity and the bulk viscosity are described as the coefficients 
of the traceless part and the trace part of $\Delta T^{ij}$, respectively. 

In the weak magnetic field regime, 3-dimensional dynamics is retained, unlike the strong 
magnetic field regime, where 3-dimensional dynamics for charged particles gets reduced to 1-dimensional dynamics and only longitudinal (along the direction of magnetic field) component of $\Delta T^{ij}$ exists. It is very important to note that, at least in the weak magnetic field limit, we do not split $\Delta T^{ij}$ into different components, rather, the effect of magnetic field enters mainly through the cyclotron frequency ($\omega_c$). According to this specific limit, we neglect the terms containing $\omega_c$ and its higher orders in the numerator. Thus, Hall-type shear and bulk viscosities are not obtained in this part of this section. In the general configuration of magnetic field, different components of aforesaid viscosities are obtained in the next part of this section. We get the spatial component of eq. \eqref{em1} 
(determined in appendix \ref{N.P. of E.M.T.}) as
\be\label{em2.qaqg1}
\nonumber\Delta T^{ij} &=& \nonumber\sum_f g_f\int\frac{d^3{\rm p}}{(2\pi)^3}\frac{\beta p^i p^j}{\omega_f}\left[2qEv_x\frac{\tau_f f_f^0\left(1-f_f^0\right)}{1+\omega_c^2\tau_f^2}+2\bar{q}Ev_x\frac{\tau_{\bar{f}}\bar{f_f}^0\left(1-\bar{f_f}^0\right)}{1+\omega_c^2\tau_{\bar{f}}^2}\right. \\ && \left.\nonumber+\left(\frac{\tau_f f_f^0\left(1-f_f^0\right)}{1+\omega_c^2\tau_f^2}+\frac{\tau_{\bar{f}}\bar{f_f}^0\left(1-\bar{f_f}^0\right)}{1+\omega_c^2\tau_{\bar{f}}^2}\right)\left\lbrace\omega_f\left(\frac{\partial P}{\partial \varepsilon}\right)-\frac{\rm p^2}{3\omega_f}\right\rbrace\partial_l u^l\right. \\ && \left.\nonumber -\left(\frac{\tau_f f_f^0\left(1-f_f^0\right)}{1+\omega_c^2\tau_f^2}+\frac{\tau_{\bar{f}}\bar{f_f}^0\left(1-\bar{f_f}^0\right)}{1+\omega_c^2\tau_{\bar{f}}^2}\right)\frac{p_kp_l}{2p_0}W_{kl}\right] \\ && \nonumber +g_g\int\frac{d^3{\rm p}}{(2\pi)^3}\frac{p^i p^j\tau_g}{\omega_g}\beta f_g^0\left(1+f_g^0\right)\left[\left\lbrace\omega_g\left(\frac{\partial P}{\partial \varepsilon}\right)-\frac{\rm p^2}{3\omega_g}\right\rbrace\partial_l u^l\right. \\ && \left.-\frac{p^kp^l}{2\omega_g}W_{kl}+p^l\left(\frac{\partial_l P}{\varepsilon+P}-\frac{\partial_l T}{T} \right)\right]
.\ee
Comparing equations \eqref{definition} and \eqref{em2.qaqg1}, we get the shear 
viscosity of a weakly magnetized hot and dense QCD matter as
\begin{eqnarray}\label{eta}
\nonumber\eta &=& \frac{\beta}{30\pi^2}\sum_f g_f \int d{\rm p}~\frac{{\rm p}^6}{\omega_f^2}\left[\frac{\tau_f}{1+\omega_c^2\tau_f^2} ~ f_f^0\left(1-f_f^0\right)+\frac{\tau_{\bar{f}}}{1+\omega_c^2\tau_{\bar{f}}^2} ~ \bar{f}_f^0\left(1-\bar{f}_f^0\right)\right] \\ && +\frac{\beta}{30\pi^2} g_g \int d{\rm p}~\frac{{\rm p}^6}{\omega_g^2} ~ \tau_g ~ f_g^0\left(1+f_g^0\right)
~.\end{eqnarray}
Similarly, the comparison between equations \eqref{definition} and 
\eqref{em2.qaqg1} gives the bulk viscosity as
\begin{eqnarray}\label{zeta.1}
\nonumber\zeta &=& \frac{1}{3}\sum_f g_f \int\frac{d^3{\rm p}}{(2\pi)^3}~\frac{{\rm p}^2}{\omega_f}\left[f_f^0\left(1-f_f^0\right)A_f+\bar{f}_f^0\left(1-\bar{f}_f^0\right)\bar{A}_f\right] \\ && +\frac{1}{3}g_g \int\frac{d^3{\rm p}}{(2\pi)^3}~\frac{{\rm p}^2}{\omega_g} ~ f_g^0\left(1+f_g^0\right)A_g
~.\end{eqnarray}
The factors $A_f$, $\bar{A}_f$ and $A_g$ in eq. \eqref{zeta.1} are 
respectively written as
\begin{eqnarray}
&&A_f=\frac{\tau_f\beta}{3\left(1+\omega_c^2\tau_f^2\right)}\left[\frac{{\rm p}^2}{\omega_f}-3\left(\frac{\partial P}{\partial \varepsilon}\right)\omega_f\right],\label{Ai} \\ 
&&\bar{A}_f=\frac{\tau_{\bar{f}}\beta}{3\left(1+\omega_c^2\tau_{\bar{f}}^2\right)}\left[\frac{{\rm p}^2}{\omega_f}-3\left(\frac{\partial P}{\partial \varepsilon}\right)\omega_f\right],\label{Ai.1} \\ 
&&A_g=\frac{\tau_g\beta}{3}\left[\frac{{\rm p}^2}{\omega_g}-3\left(\frac{\partial P}{\partial \varepsilon}\right)\omega_g\right]\label{Ag}
.\end{eqnarray}
The calculation of viscosity requires nonzero velocity gradient. But there exist different frames to define velocity $u^\mu$, for example, $u^\mu$ denotes the velocity of baryon number flow in the Eckart frame, whereas it denotes the velocity of energy flow in the Landau-Lifshitz frame. Therefore the freedom to choose a specific frame creates arbitrariness. To avoid this arbitrariness, one needs the ``condition of fit'', {\em i.e.} if one chooses the Landau-Lifshitz frame, then the condition of fit in the local rest frame demands the ``$00$'' component of the dissipative part of the energy-momentum tensor to 
be zero ($\Delta T^{00}=0$). In order to satisfy this Landau-Lifshitz condition, the factors $A_f$, $\bar{A}_f$ and $A_g$ should be replaced as $A_f\rightarrow A_f^\prime=A_f-b_f\omega_f$, $\bar{A}_f\rightarrow \bar{A}_f^\prime=\bar{A}_f-\bar{b}_f\omega_f$ and $A_g\rightarrow A_g^\prime=A_g-b_g\omega_g$. The Landau-Lifshitz conditions for $A_f$, $\bar{A}_f$ and $A_g$ are respectively given by
\begin{eqnarray}
&&\sum_f g_f\int\frac{d^3{\rm p}}{(2\pi)^3}~\omega_f f_f^0(1-f_f^0)\left(A_f-b_f\omega_f\right)=0 \label{A_i} ~,~ \\ 
&&\sum_f g_f\int\frac{d^3{\rm p}}{(2\pi)^3}~\omega_f \bar{f}_f^0(1-\bar{f}_f^0)\left(\bar{A}_f-\bar{b}_f\omega_f\right)=0 \label{A_i.1} ~,~ \\ 
&&g_g\int\frac{d^3{\rm p}}{(2\pi)^3}~\omega_g f_g^0(1+f_g^0)\left(A_g-b_g\omega_g\right)=0 \label{A_g}
~.\end{eqnarray}
The quantities $b_f$, $\bar{b}_f$ and $b_g$ are arbitrary constants and are associated with the particle and energy conservations for a thermal medium having asymmetry between the numbers of particles and antiparticles \cite{Chakraborty:PRC83'2011}. These quantities can be obtained by solving equations \eqref{A_i}, \eqref{A_i.1} and \eqref{A_g}. After substituting $A_f\rightarrow A_f^\prime$, $\bar{A}_f\rightarrow \bar{A}_f^\prime$ and 
$A_g\rightarrow A_g^\prime$ in eq. \eqref{zeta.1} and simplifying, we get the bulk 
viscosity of a weakly magnetized hot and dense QCD matter as
\begin{eqnarray}\label{zeta}
\nonumber\zeta &=& \frac{\beta}{18\pi^2}\sum_f g_f \int d{\rm p}~{\rm p}^2\left[\frac{{\rm p}^2}{\omega_f}-3\left(\frac{\partial P}{\partial \varepsilon}\right)\omega_f\right]^2\left[\frac{\tau_f}{1+\omega_c^2\tau_f^2} ~ f_f^0\left(1-f_f^0\right)+\frac{\tau_{\bar{f}}}{1+\omega_c^2\tau_{\bar{f}}^2} ~ \bar{f}_f^0\left(1-\bar{f}_f^0\right)\right] \\ && +\frac{\beta}{18\pi^2}g_g\int d{\rm p}~{\rm p}^2\left[\frac{{\rm p}^2}{\omega_g}-3\left(\frac{\partial P}{\partial \varepsilon}\right)\omega_g\right]^2 \tau_g ~ f_g^0\left(1+f_g^0\right)
~.\end{eqnarray}
In this part, we have obtained the shear and bulk viscosities using the ansatz method in the weak magnetic field limit. In the next part, we are going to determine different components of shear and bulk viscosities in the general configuration of magnetic field. 

\nd\underline{\bf Momentum transport coefficients in the general configuration of magnetic field}
In the presence of an arbitrary magnetic field, the infinitesimal change in the distribution function of charged particles (quarks and antiquarks) is written as
\be\label{small d.f.}
\delta f=\sum_{l=0}^4C_lY_{mn}^lv_mv_n
.\ee
The spatial component of the nonequilibrium part of the energy-momentum tensor is written as
\be\label{d.p.(1)}
\Delta T_{ij}=\sum_{l=0}^4\eta_lY_{ij}^l
,\ee
where $\eta_0$, $\eta_1$, $\eta_2$, $\eta_3$ and $\eta_4$ denote five shear viscosity coefficients. For the calculation of the viscosities, it is sufficient to take only spatial component of the nonequilibrium part of the energy-momentum tensor. In the above tensor, we have excluded the bulk viscosity part to determine the shear viscosity coefficients. In terms of the infinitesimal change in the particle distribution function, $\Delta T_{ij}$ has the following form, 
\be\label{d.p.(2)}
\Delta T_{ij}=\sum_fg_f\int\frac{d^3{\rm p}}{(2\pi)^3}\omega_fv_iv_j\delta f
.\ee
Substituting the value of $\delta f$ \eqref{small d.f.} in eq. \eqref{d.p.(2)} and then simplifying, we get
\be\label{d.p.(3)}
\Delta T_{ij}=\frac{1}{15}\sum_fg_f\int\frac{d^3{\rm p}}{(2\pi)^3}\omega_fv^4\left(\delta_{ij}\delta_{mn}+\delta_{im}\delta_{jn}
+\delta_{in}\delta_{jm}\right)\sum_{l=0}^4C_lY_{mn}^l
.\ee
In the above equations, $Y_{ij}^0$, $Y_{ij}^1$, $Y_{ij}^2$, $Y_{ij}^3$ and $Y_{ij}^4$ are respectively expressed \cite{Lifshitz:BOOK'1981,Tuchin:JPG39'2012} as
\begin{eqnarray}\label{Form (1)}
Y_{ij}^0 &=& \left(3 b_i b_j-\delta_{ij}\right)\left(b_k b_l V_{kl}-\frac{1}{3}\nabla \cdot V\right), \\ 
Y_{ij}^1 &=& 2V_{ij}+\left(b_i b_j - \delta_{ij}\right)\nabla\cdot\mathbf{V} + \delta_{ij}V_{kl}b_k b_l-2V_{ik}b_k b_j-2V_{jk}b_k b_i+b_i b_j V_{kl}b_k b_l,\label{Form (2)} \\ 
Y_{ij}^2 &=& 2V_{ik}b_k b_j+2V_{jk}b_k b_i-4b_i b_j V_{kl}b_k b_l,\label{Form (3)} \\ 
Y_{ij}^3 &=& V_{ik}b_{jk}+V_{jk}b_{ik}-V_{kl}b_{ik}b_j b_l-
V_{kl}b_{jk}b_i b_l,\label{Form (4)} \\
Y_{ij}^4 &=& 2V_{kl}b_{ik}b_j b_l+2V_{kl}b_{jk}b_i b_l\label{Form (5)} 
,\end{eqnarray}
where $b_{ij}=\epsilon_{ijk}b_k$ and $V_{ij}=\frac{1}{2}\left(\frac{\partial V_i}{\partial x_j}+\frac{\partial V_j}{\partial x_i}\right)$, with $V_i$ and $b_i=\frac{\bf B}{\rm B}$ denoting the fluid velocity and the unit vector along the direction of magnetic field, respectively. Imposing the condition, $\nabla\cdot\mathbf{V}=0$, and using the relations, such as $V_{ij}b_ib_j=0$, $b_{ij}v_iv_j=0$, $b_ib_i=1$, $b_{ij}b_i=0$ and $b_{ij}b_j=0$, we determine $\eta_1$, $\eta_2$, $\eta_3$ and $\eta_4$. On the other hand, $\eta_0$ remains the same as in the absence of magnetic field and is given by
\be\label{eta (0)}
\eta_0=\frac{\beta}{30\pi^2}\sum_f g_f \int d{\rm p}~\frac{{\rm p}^6}{\omega_f^2}\left[\tau_f f_f^0\left(1-f_f^0\right)+\tau_{\bar{f}} \bar{f}_f^0\left(1-\bar{f}_f^0\right)\right]
.\ee
Comparing eq. \eqref{d.p.(1)} and eq. \eqref{d.p.(3)}, and requiring the consistency of both these equations, we have
\be\label{eta.1}
&&\eta_1=\frac{2}{15}\sum_f g_f\int\frac{d^3{\rm p}}{(2\pi)^3}\omega_fv^4C_1, \\ 
&&\eta_2=\frac{2}{15}\sum_f g_f\int\frac{d^3{\rm p}}{(2\pi)^3}\omega_fv^4C_2, \label{eta.2} \\ 
&&\eta_3=-\frac{2}{15}\sum_f g_f\int\frac{d^3{\rm p}}{(2\pi)^3}\omega_fv^4C_3, \label{eta.3} \\ 
&&\eta_4=-\frac{2}{15}\sum_f g_f\int\frac{d^3{\rm p}}{(2\pi)^3}\omega_fv^4C_4 \label{eta.4}
.\ee
The factors $C_1$, $C_2$, $C_3$ and $C_4$ can be calculated by using the relativistic Boltzmann transport equation in the relaxation time approximation at finite magnetic field and chemical potential \eqref{R.B.T.E.}. To proceed for the calculation, we take only the spatial components in eq. \eqref{R.B.T.E.} and keep only the magnetic field part in the Lorentz force. Then, we split $f$ as $f=f_0+\delta f$ in the left hand side of eq. \eqref{R.B.T.E.} and keep only $f_0$. In doing so, the second term will vanish due to the appearance of the expression $q\left(\mathbf{v}\times\mathbf{B}\right)\cdot\frac{\partial f_0}{\partial \mathbf{p}}=-q\beta\left[\left(\mathbf{v}\times\mathbf{B}\right)\cdot\mathbf{v}\right]f_0\left(1-f_0\right)=0$. So, in order to keep the magnetic field dependence, we need to keep $\delta f$ in the second term. Thus, eq. \eqref{R.B.T.E.} gets simplified into 
\be\label{R.B.T.E.(1)}
\frac{p_i}{\omega_f}\frac{\partial f_0}{\partial x_i}-\frac{qB}{\omega_f}b_{ij}v_j\frac{\partial (\delta f)}{\partial v_i}=-\frac{\delta f}{\tau_f}
,\ee
where $\frac{p_i}{\omega_f}\frac{\partial f_0}{\partial x_i}=-\beta\omega_fv_iv_jV_{ij}f_0\left(1-f_0\right)$ and the value of $\delta f$ is given in equation \eqref{small d.f.}. Now, eq. \eqref{R.B.T.E.(1)} becomes
\be\label{R.B.T.E.(2)}
\nonumber\beta\omega_fV_{ij}v_iv_jf_0\left(1-f_0\right) &=& -\omega_cb_{ij}v_j\frac{\partial}{\partial v_i}\left(\sum_{l=0}^4C_lY_{mn}^lv_mv_n\right)+\frac{\sum_{l=0}^4C_lY_{mn}^lv_mv_n}{\tau_f} \\ &=& -2\omega_cb_{ij}v_j\left(\sum_{l=0}^4C_lY_{im}^lv_m\right)
+\frac{\sum_{l=0}^4C_lY_{mn}^lv_mv_n}{\tau_f}
.\ee
Using the relations $\nabla\cdot\mathbf{V}=0$, $V_{ij}b_ib_j=0$, $b_{ij}v_iv_j=0$, $b_ib_i=1$, $b_{ij}b_i=0$ and $b_{ij}b_j=0$ in above equation, and then comparing the same tensor structures on both sides of eq. \eqref{R.B.T.E.(2)}, $C_1$, $C_2$, $C_3$ and $C_4$ can be obtained (in appendix \ref{C1C2C3C4}) as
\be
&&C_1=\frac{\beta\omega_f\tau_ff_0\left(1-f_0\right)}{2\left(1+4\omega_c^2\tau_f^2\right)}, \\ 
&&C_2=\frac{\beta\omega_f\tau_ff_0\left(1-f_0\right)}{2\left(1+\omega_c^2\tau_f^2\right)}, \\ 
&&C_3=-\frac{\beta\omega_f\omega_c\tau_f^2f_0\left(1-f_0\right)}{\left(1+4\omega_c^2\tau_f^2\right)}, \\ 
&&C_4=-\frac{\beta\omega_f\omega_c\tau_f^2f_0\left(1-f_0\right)}{2\left(1+\omega_c^2\tau_f^2\right)}
.\ee
Substituting the values of $C_1$, $C_2$, $C_3$ and $C_4$ in equations \eqref{eta.1}, \eqref{eta.2}, \eqref{eta.3} and \eqref{eta.4} and then simplifying, we get $\eta_1$, $\eta_2$, $\eta_3$ and $\eta_4$ respectively as
\be\label{eta (1)}
&&\eta_1=\frac{\beta}{30\pi^2}\sum_f g_f \int d{\rm p}~\frac{{\rm p}^6}{\omega_f^2}\left[\frac{\tau_f}{1+4\omega_c^2\tau_f^2} ~ f_f^0\left(1-f_f^0\right)+\frac{\tau_{\bar{f}}}{1+4\omega_c^2\tau_{\bar{f}}^2} ~ \bar{f}_f^0\left(1-\bar{f}_f^0\right)\right], \\ 
&&\eta_2=\frac{\beta}{30\pi^2}\sum_f g_f \int d{\rm p}~\frac{{\rm p}^6}{\omega_f^2}\left[\frac{\tau_f}{1+\omega_c^2\tau_f^2} ~ f_f^0\left(1-f_f^0\right)+\frac{\tau_{\bar{f}}}{1+\omega_c^2\tau_{\bar{f}}^2} ~ \bar{f}_f^0\left(1-\bar{f}_f^0\right)\right], \label{eta (2)} \\ 
&&\eta_3=\frac{\beta}{30\pi^2}\sum_f g_f \int d{\rm p}~\frac{{\rm p}^6}{\omega_f^2}\left[\frac{2\omega_c\tau_f^2}{1+4\omega_c^2\tau_f^2}f_f^0\left(1-f_f^0\right)+\frac{2\omega_c\tau_{\bar{f}}^2}{1+4\omega_c^2\tau_{\bar{f}}^2}\bar{f}_f^0\left(1-\bar{f}_f^0\right)\right], \label{eta (3)} \\ 
&&\eta_4=\frac{\beta}{30\pi^2}\sum_f g_f \int d{\rm p}~\frac{{\rm p}^6}{\omega_f^2}\left[\frac{\omega_c\tau_f^2}{1+\omega_c^2\tau_f^2}f_f^0\left(1-f_f^0\right)+\frac{\omega_c\tau_{\bar{f}}^2}{1+\omega_c^2\tau_{\bar{f}}^2}\bar{f}_f^0\left(1-\bar{f}_f^0\right)\right] \label{eta (4)}
.\ee
Neglecting the factor 4 in the denominator of eq. \eqref{eta (1)}, one can find that $\eta_1=\eta_2=\eta$ (charged particle part), where $\eta$ is given in eq. \eqref{eta}. In the above description of different shear viscosity coefficients, the gluon part of the shear viscosity has been excluded, because magnetic field has almost no effect on the electrically neutral gluons, thus, this part of the viscosity does not split into different components in the presence of magnetic field. So, one can add the gluon part to the charged particle part to get the total shear viscosity of the hot medium of quarks, antiquarks and gluons like in eq. \eqref{eta}. Now, excluding the shear viscosity part and including only the bulk viscosity part, $\Delta T_{ij}$ is expressed \cite{Lifshitz:BOOK'1981} as
\be\label{d.p.(4)}
\Delta T_{ij}=\zeta_0 \delta_{ij} \nabla \cdot \mathbf V+\zeta_1\left(\delta_{ij}V_{kl}b_k b_l+b_i b_j\nabla\cdot\mathbf{V}\right)
.\ee
Thus, there also exist two different bulk viscosity coefficients in the presence of an arbitrary magnetic field, such as $\zeta_0$ and $\zeta_1$. The volume or bulk viscosity coefficient $\zeta_0$ remains the same as in the absence of magnetic field and is given by
\be\label{zeta (0)}
\zeta_0=\frac{\beta}{18\pi^2}\sum_f g_f \int d{\rm p}~{\rm p}^2\left[\frac{{\rm p}^2}{\omega_f}-3\left(\frac{\partial P}{\partial \varepsilon}\right)\omega_f\right]^2\left[\tau_f f_f^0\left(1-f_f^0\right)+\tau_{\bar{f}} \bar{f}_f^0\left(1-\bar{f}_f^0\right)\right]
.\ee
On the other hand, $\zeta_1$ which is the cross effect between the ordinary and volume viscosities vanishes for a plasma (for details, please see ref. \cite{Lifshitz:BOOK'1981}). Thus, to see the magnetic field-dependence, we use the bulk viscosity obtained through the ansatz method at weak magnetic field limit in the first part of this section \eqref{zeta}. 

The aforementioned transport properties are studied considering the quasiparticle model (QPM) of QGP medium. In quasiparticle models \cite{Peshier:PLB337'1994,Gorenstein:PRD52'1995,Bluhm:EPJC49'2007,Berrehrah:PRC89'2014}, QGP is described as a system of massive noninteracting quasiparticles and the mass of the quasiparticle arises due to the interactions of quarks and gluons with the thermal 
medium. In the kinetic theory approach with the quasiparticle model 
description, the interactions among partons have been considered to be 
contained only in their quasiparticle masses. Quasiparticle masses of particles have been derived from the hard thermal loop (HTL) perturbation theory at high temperatures \cite{Braaten:PRD45'1992,Peshier:PRD66'2002}. We note that the estimation of the quasiparticle model depends on the requirement of thermodynamic consistency, which has already been tested in notable works like \cite{Peshier:PLB337'1994,Gorenstein:PRD52'1995,Peshier:PRD66'2002}. It 
assumes that the deconfined quarks and gluons remain the relevant degrees of 
freedom even in the quasiparticle model, which is a justified assumption for high temperatures $T>T_c$ and for small chemical potentials $\mu<2T_c$, because this 
model reproduces the leading-order perturbative results and in addition, it 
represents a thermodynamically consistent effective resummation of the 
leading-order thermal contributions. For the thermodynamic 
consistency, there are some conditions which need to be satisfied, for example, the 
derivative of pressure with respect to the square of quasiparticle mass requires to 
vanish. At high temperatures, the thermodynamic consistency can be fulfilled, because the thermodynamic quantities can be perturbatively expanded in powers of coupling $g$ 
and the full expressions represent a thermodynamically consistent 
resummation of terms of all orders in coupling $g$ \cite{Peshier:PRD66'2002}. Thus, the coupling must be very small and this is unambiguously satisfied in high temperature 
QGP phase \cite{Peshier:PLB337'1994,Peshier:PRD66'2002}. 

The quasiparticle model was successfully used to study the equation of state for the partonic medium \cite{Goloviznin:ZPC57'1993,Peshier:PRD54'1996}. This model had also been studied in different approaches, such as the Nambu-Jona-Lasinio (NJL) and Polyakov NJL based quasiparticle models \cite{Fukushima:PLB591'2004,Ghosh:PRD73'2006,Abuki:PLB676'2009}, quasiparticle model in a strong magnetic field \cite{Rath:PRD102'2020,Rath:EPJC81'2021}, quasiparticle model with Gribov-Zwanziger quantization \cite{Su:PRL114'2015,Florkowski:PRC94'2016}, thermodynamically consistent quasiparticle model \cite{Bannur:JHEP0709'2007,Bannur:PRC75'2007} etc. In a hot and dense medium, the thermal mass (squared) of quark is given \cite{Braaten:PRD45'1992,Peshier:PRD66'2002} by
\be\label{Q.P.M.}
m_{fT}^2=\frac{g^2T^2}{6}\left(1+\frac{\mu_f^2}{\pi^2T^2}\right)
.\ee
In the similar environment, the thermal mass (squared) of gluon is given \cite{Peshier:PRD66'2002,Blaizot:PRD72'2005,Berrehrah:PRC89'2014} by
\be\label{Q.P.M.(Gluon mass)}
m_{gT}^2=\frac{g^2T^2}{6}\left(N_c+\frac{N_f}{2}+\frac{3}{2\pi^2T^2}\sum_f\mu_f^2\right)
.\ee
In the above equations, $g^2=4\pi\alpha_s$, where $\alpha_s$ denotes the one-loop strong running coupling at finite temperature, chemical potential and weak magnetic field, and is expressed \cite{Ayala:PRD98'2018} as
\be\label{R.C.}
\alpha_s\left(\Lambda^2, eB\right)=\frac{\alpha_s\left(\Lambda^2\right)}{1+b_1\alpha_s\left(\Lambda^2\right)\ln\left(\frac{\Lambda^2}{\Lambda^2+eB}\right)}
.\ee
Here $\alpha_s\left(\Lambda^2\right)$ is the one-loop strong running 
coupling in the absence of magnetic field, which is given by
\be\label{R.C.1}
\alpha_s\left(\Lambda^2\right)=\frac{1}{b_1\ln\left(\frac{\Lambda^2}{\Lambda_{\rm\overline{MS}}^2}\right)}
,\ee
with $b_1=\frac{11N_c-2N_f}{12\pi}$, $\Lambda_{\rm\overline{MS}}=0.176$ GeV 
and $\Lambda=2\pi\sqrt{T^2+\mu_f^2/\pi^2}$ for electrically charged 
particles (quarks and antiquarks) and $\Lambda=2 \pi T$ for gluons. The 
chemical potentials for all flavors have been kept the same, {\em i.e.} $\mu_f=\mu$. 

\section{Results and discussions}
\begin{figure}[]
\begin{center}
\begin{tabular}{c c}
\includegraphics[width=6.86cm]{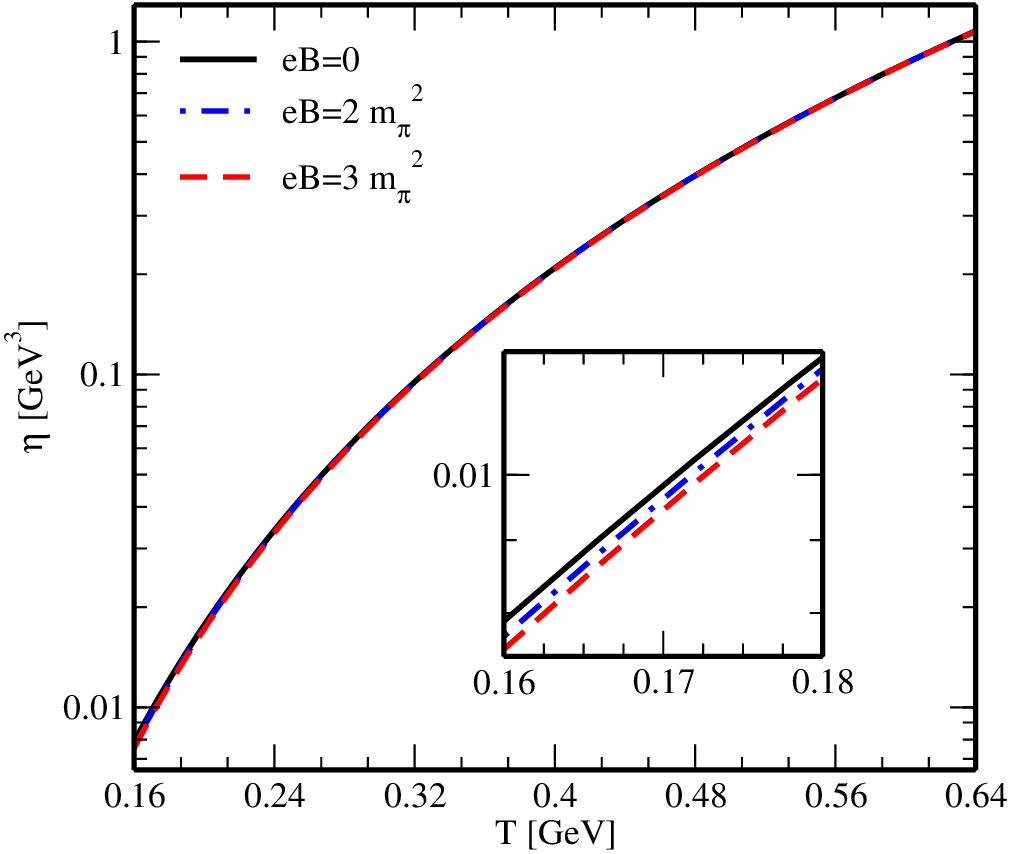}&
\hspace{0.74 cm}
\includegraphics[width=6.86cm]{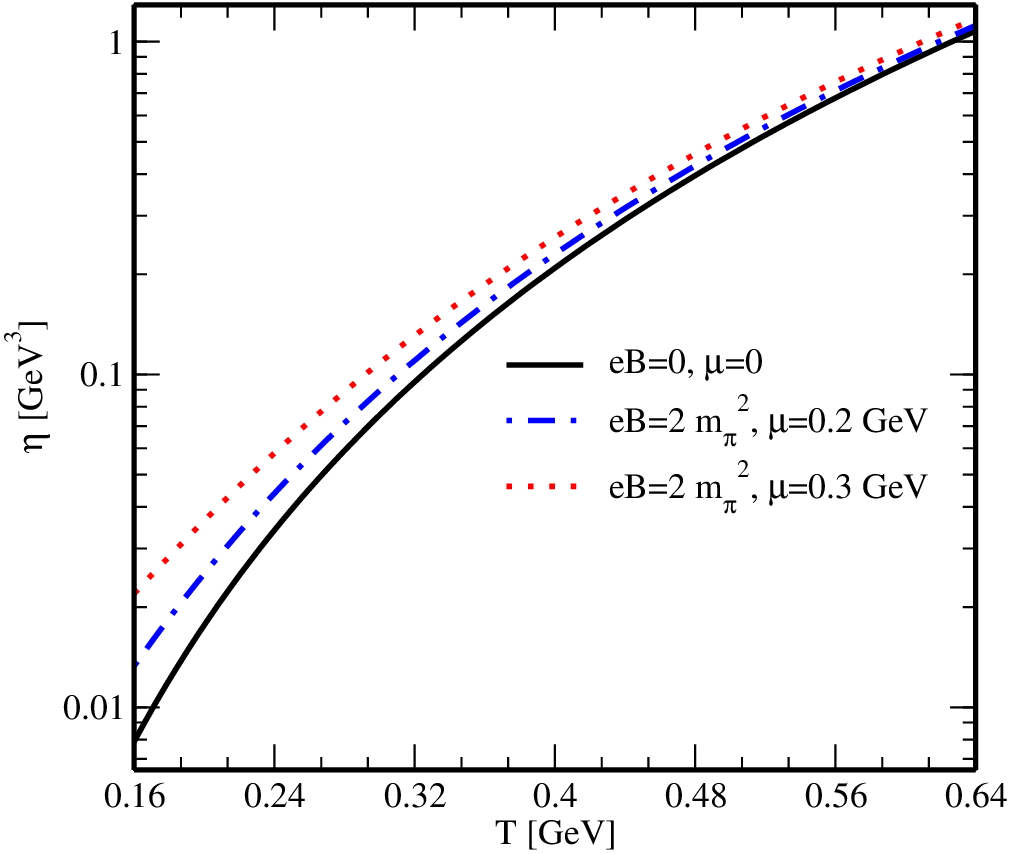} \\
a & b
\end{tabular}
\caption{The variation of the shear viscosity with temperature 
(a) in the presence of weak magnetic field and (b) in the 
presence of finite chemical potential.}\label{Fig.1}
\end{center}
\end{figure}

\begin{figure}[]
\begin{center}
\begin{tabular}{c c}
\includegraphics[width=6.86cm]{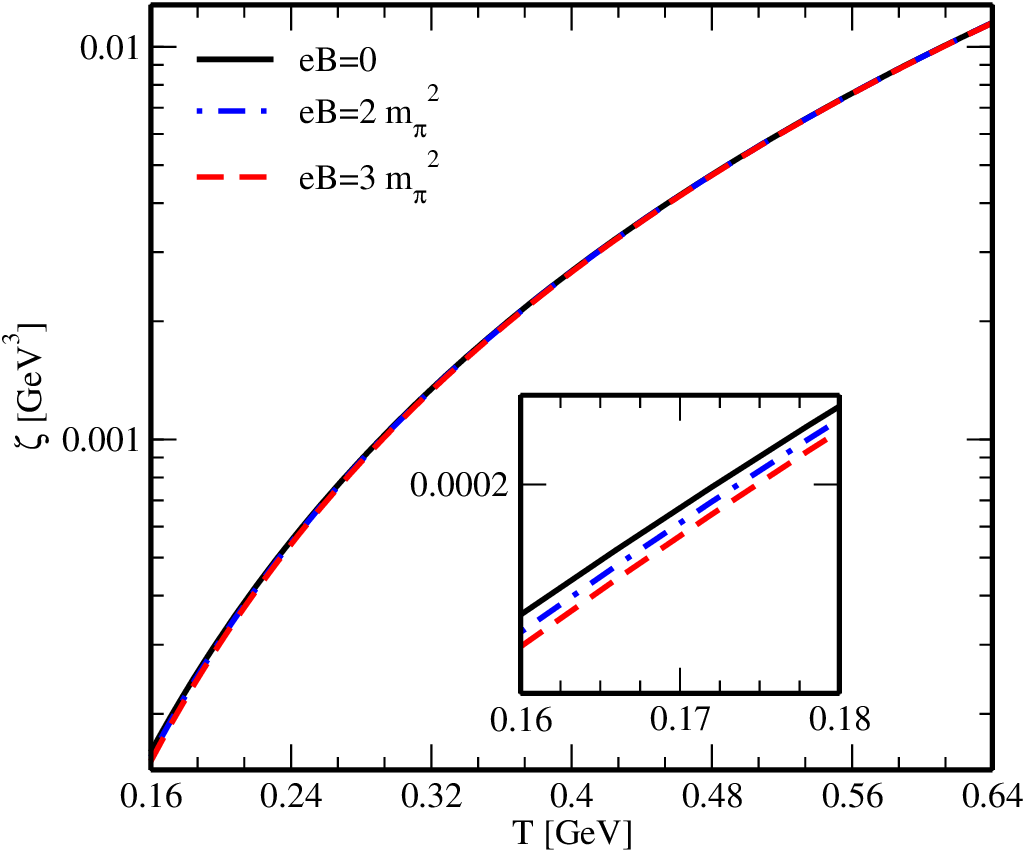}&
\hspace{0.74 cm}
\includegraphics[width=6.86cm]{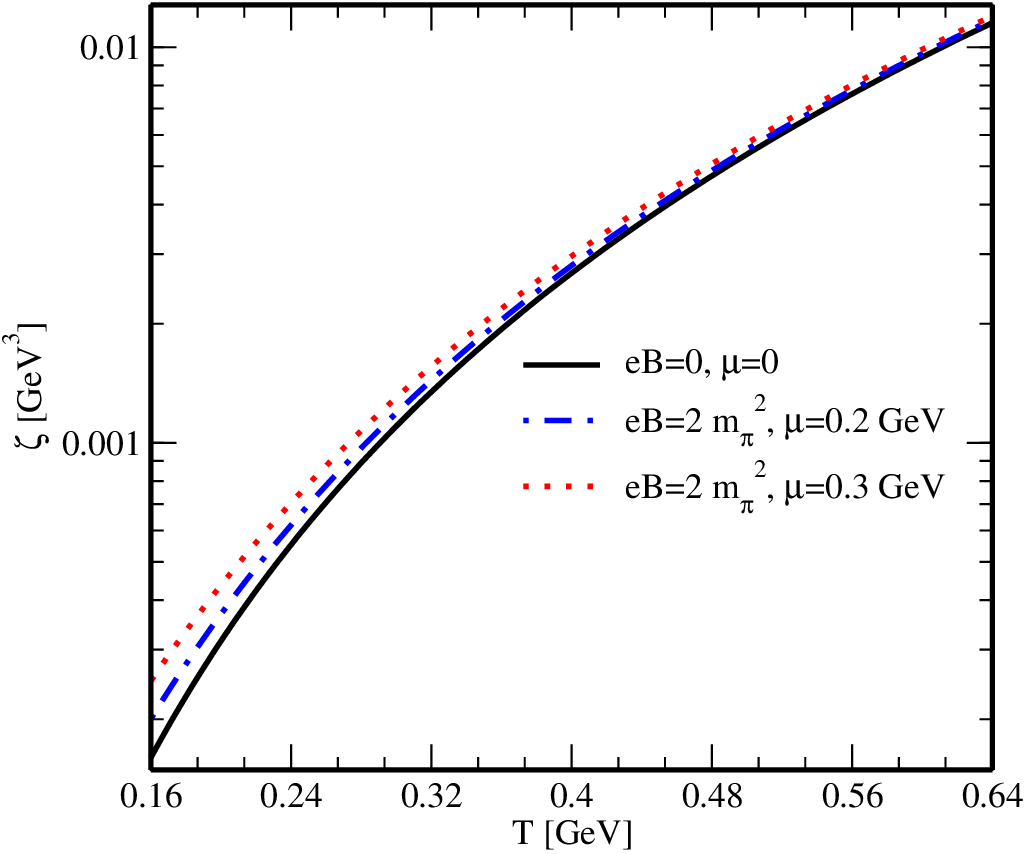} \\
a & b
\end{tabular}
\caption{The variation of the bulk viscosity with temperature 
(a) in the presence of weak magnetic field and (b) in the 
presence of finite chemical potential.}\label{Fig.2}
\end{center}
\end{figure}

\begin{figure}[]
\begin{center}
\begin{tabular}{c c}
\includegraphics[width=6.86cm]{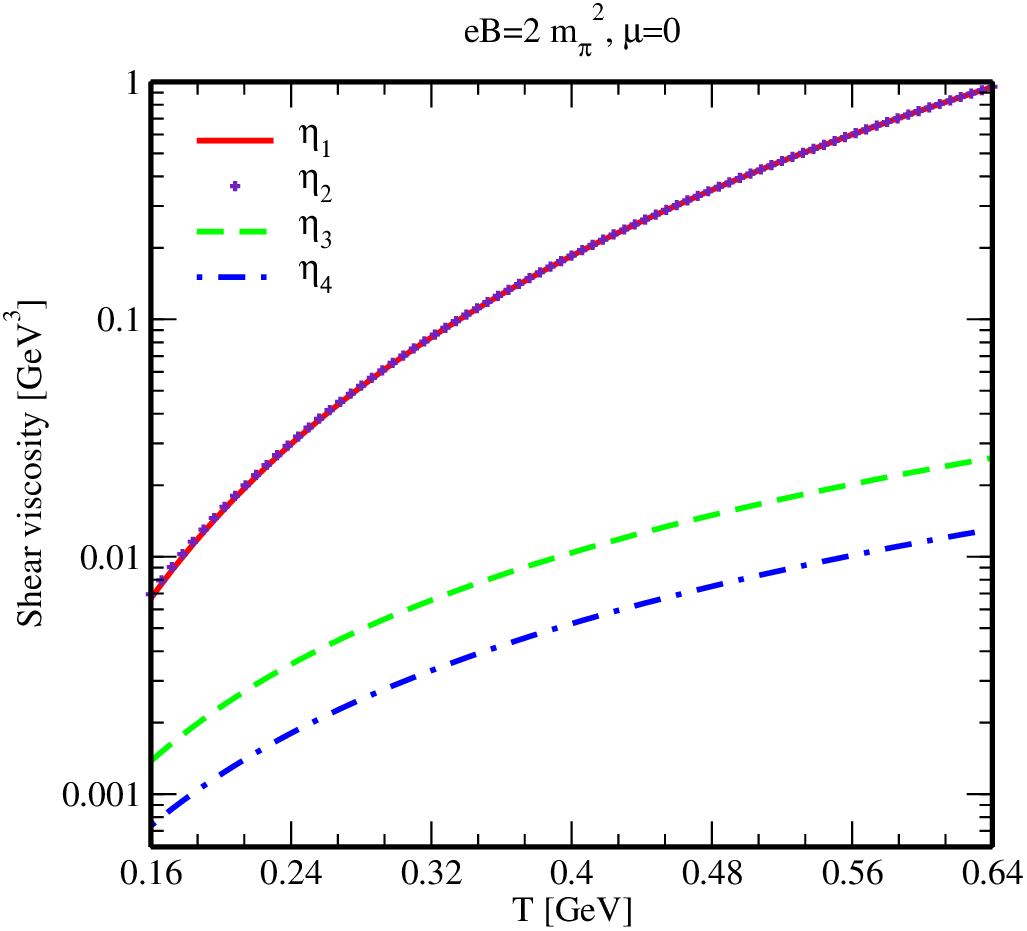}&
\hspace{0.74 cm}
\includegraphics[width=6.86cm]{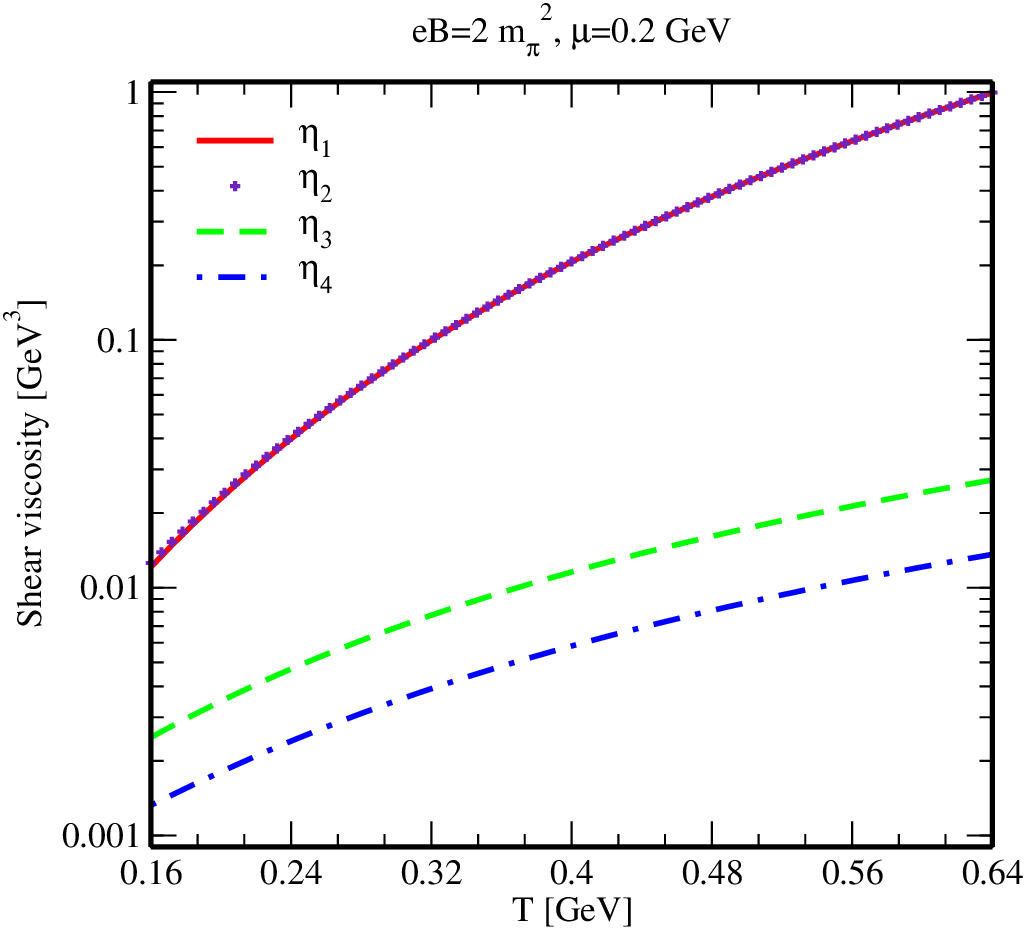} \\
a & b
\end{tabular}
\caption{Variations of the four shear viscosity coefficients with temperature 
(a) in the presence of weak magnetic field and (b) in the presence of finite 
chemical potential.}\label{Fig.1234}
\end{center}
\end{figure}

Figures \ref{Fig.1} and \ref{Fig.2} show the temperature dependence of shear ($\eta$) and bulk ($\zeta$) viscosities in the presence of weak magnetic field and finite chemical potential, respectively. It can be seen from these figures that the influence of weak magnetic field on $\eta$ and $\zeta$ is less pronounced than the influence of chemical potential. Compared to the thermal medium at $eB=0$, $\mu=0$, the decrease of $\eta$ and $\zeta$ due to the weak magnetic field is meagre, contrary to their discernible increase due to the finite chemical potential. These effects of weak magnetic field and chemical potential on shear and bulk viscosities are more conspicuous at low temperatures. Thus, the reduction in $\eta$ leads to a decrease in the momentum transport in the presence of weak magnetic field, whereas the finite chemical potential creates favorable condition for the momentum transport in hot QCD matter and it becomes easy for a particle to carry momentum over great distances. It can also be inferred that their effects on the momentum transport get suppressed at higher temperatures. Further, the reduction in $\zeta$ in weak magnetic field regime explains small fluctuations in the pressure, contrary to large fluctuations at finite chemical potential. At finite magnetic field, the magnetic catalysis phenomenon enhances the dynamical symmetry breaking, thus triggering the binding of oppositely charged particles. It results in a stronger interaction between the constituents of the medium, which thus reduces the viscosities. In addition, with the magnetic field, cyclotron frequency increases and particle distributions decrease, which also give a decreasing effect to the viscosities. But, in the weak magnetic field limit this decrease is meagre, which can be understood from the fact that, in this limit, the magnetic field is the weak energy scale and the temperature is the strong energy scale. So, at high temperature phase, the effects of weak magnetic field on the abovementioned quantities and phenomenon are less pronounced. Thus, the viscosities have a negligible dependence on the magnetic field. Throughout the temperature range, the shear viscosity remains nearly two orders of magnitude larger than the bulk viscosity. Thus, the momentum transfer across the layer exceeds the momentum transfer along the layer. The dominance of shear viscosity over bulk viscosity also describes that the change in shape at constant volume is dominant as compared to the change in volume at constant shape. 

The enhancement of shear viscosity at finite chemical potential also supports the reduction of elliptic flow in the similar regime, which can be understood as follows. We know that $v_2$ measures the flow anisotropy in the azimuthal plane. The shear viscosity being a result of frictional force and the frictional force being proportional to the flow velocity have noticeably large effects on the fast-moving particles in the collision plane. Thus anisotropy gets reduced, resulting a decrease in $v_2$ at finite chemical potential. Although the bulk viscosity is very small, but the emergence of finite chemical potential tends to enhance its magnitude. It thus explains that the chemical potential supports the deviation of the strongly interacting matter from conformality. 

For the comparison, we have plotted four shear viscosity coefficients, $\eta_1$, $\eta_2$, $\eta_3$ and $\eta_4$ as functions of temperature at weak magnetic field and finite chemical potential in figure \ref{Fig.1234}. One can see that, $\eta_1$ and $\eta_2$ are almost indistinguishable, whereas $\eta_3$ and $\eta_4$ are distinguishable. This is expected, because the appearance of factor 4 in the denominator does not affect much, so $\eta_1$ \eqref{eta (1)} is almost equal to $\eta_2$ \eqref{eta (2)}, whereas the appearance of factor 2 in the numerator does affect noticeably, so the difference between $\eta_3$ \eqref{eta (3)} and $\eta_4$ \eqref{eta (4)} is conspicuous. Both $\eta_3$ and $\eta_4$ directly depend on magnetic field through the cyclotron frequency $\omega_c$ (can be seen in equations \eqref{eta (3)} and \eqref{eta (4)}) and hence called Hall-type shear viscosity coefficients. These Hall-type shear viscosity coefficients $\eta_3$ and $\eta_4$ are found to be much smaller than the shear viscosity coefficients $\eta_1$ and $\eta_2$, which explains that, $\eta_2(\approx\eta_1)$ is the dominant shear viscosity coefficient. One can also notice that, $\eta_2(\approx\eta_1)$ is exactly equal to the charged particle part of $\eta$ obtained using the ansatz method in the first part of section 2, {\em i.e.} $\eta_2(\approx\eta_1)=\eta$ (charged particle part). As compared to the $\mu=0$ case (figure \ref{Fig.1234}a), these coefficients get increased at finite chemical potential (figure \ref{Fig.1234}b). 

\section{Applications}
In this section, we are going to study the effects of weak magnetic field and finite 
chemical potential on the Prandtl number, the Reynolds number, the ratio of shear 
viscosity to entropy density, $\eta/s$ and the ratio of bulk viscosity to entropy 
density, $\zeta/s$. 

\subsection{Prandtl number}
The momentum diffusion and the thermal diffusion are not completely independent, 
rather, they are related through the Prandtl number (Pr) as
\begin{equation}\label{Pl}
{\rm Pr}=\frac{\eta/\rho}{\kappa/C_p}
~,\end{equation}
where $C_p$ represents the specific heat at constant 
pressure, $\rho$ is the mass density and $\kappa$ 
denotes the thermal conductivity. The Prandtl number 
is important to understand the effects of momentum 
diffusion and thermal diffusion on the sound attenuation 
in a medium. For Pr$<$1, the dominance of thermal 
diffusion over momentum diffusion in the sound attenuation 
is implied, unlike the case where Pr$>$1. The estimation 
of the Prandtl number is carried out in a weak magnetic field, 
using the expression of the thermal conductivity in the 
similar environment (written in appendix \ref{A.T.C.}) from 
our recent work \cite{Rath:2112.11802}. $C_p$ and $\rho$ are 
calculated from the energy-momentum tensor ($C_p=\partial (u_\mu T^{\mu\nu}u_\nu-\Delta_{\mu\nu}T^{\mu\nu}/3)/\partial T$, 
with $\Delta_{\mu\nu}=g_{\mu\nu}-u_\mu u_\nu$) and the particle flow four-vector 
($\rho=\sum_{f,\bar{f},g} m_{f,\bar{f},g}u_\mu N^\mu$, with 
$m_{f,\bar{f},g}$ denoting the quasiparticle mass), respectively. 

\begin{figure}[]
\begin{center}
\begin{tabular}{c c}
\includegraphics[width=6.86cm]{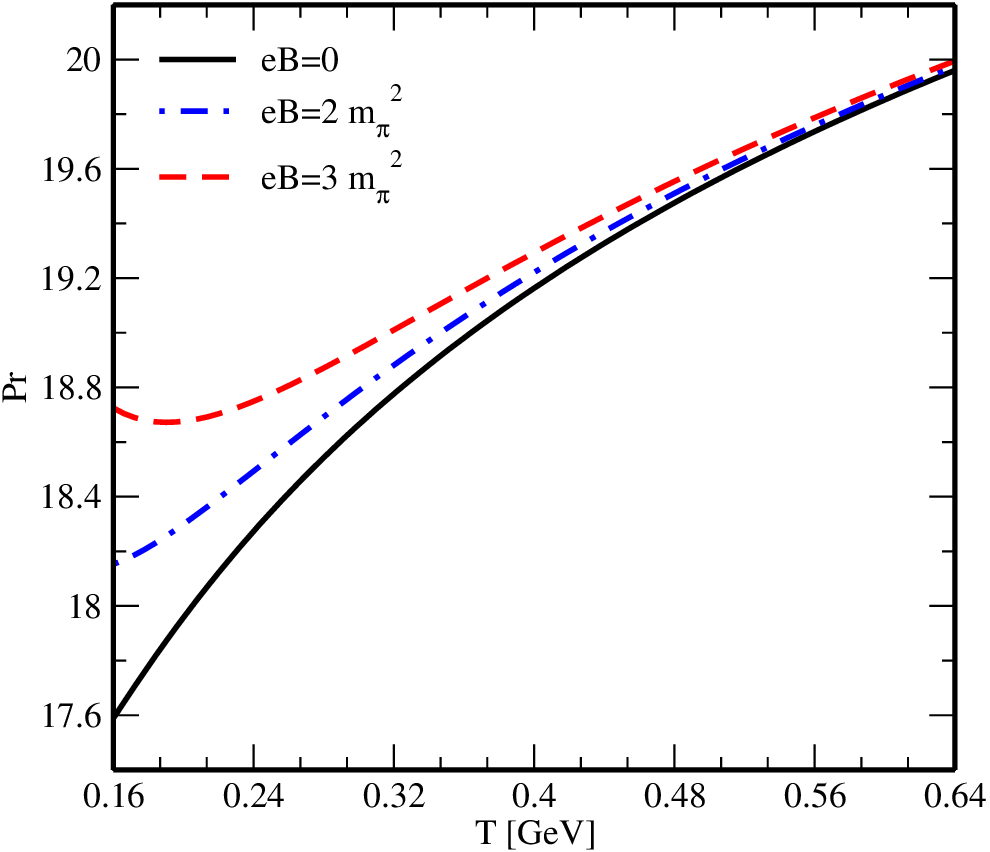}&
\hspace{0.74 cm}
\includegraphics[width=6.86cm]{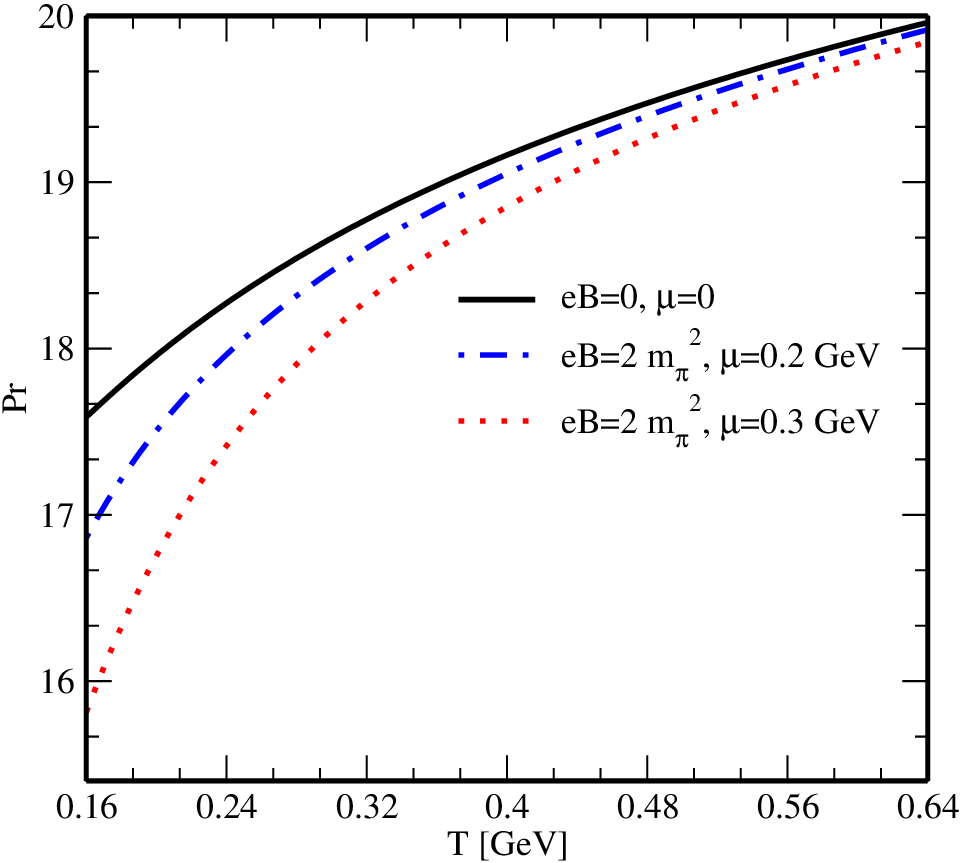} \\
a & b
\end{tabular}
\caption{The variation of the Prandtl number with temperature (a) in the 
presence of weak magnetic field and (b) in the presence of finite 
chemical potential.}\label{pl.1}
\end{center}
\end{figure}

\begin{figure}[]
\begin{center}
\begin{tabular}{c c}
\includegraphics[width=6.86cm]{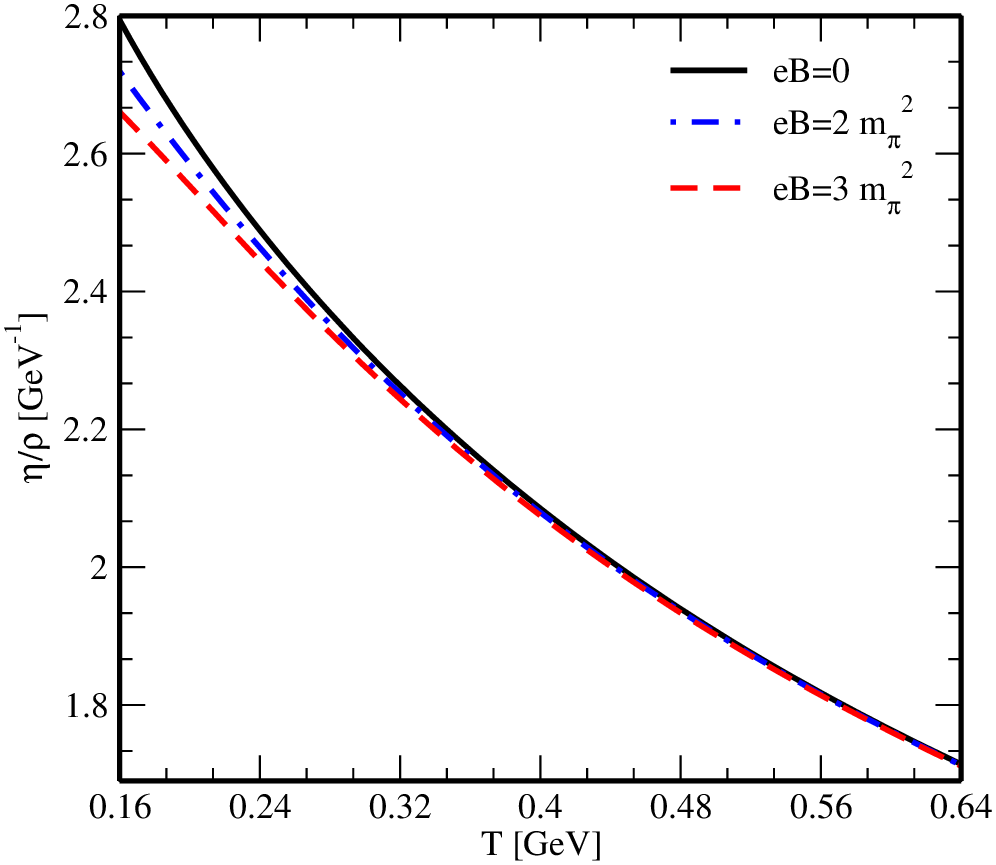}&
\hspace{0.74 cm}
\includegraphics[width=6.86cm]{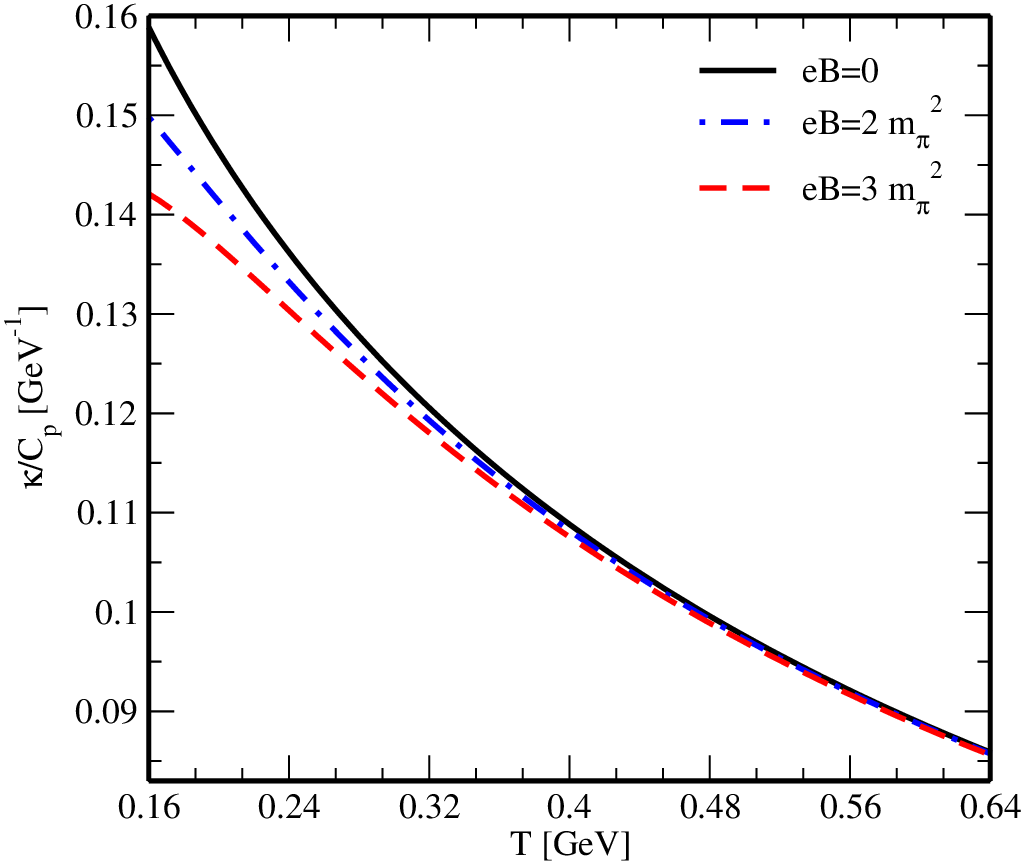} \\
a & b
\end{tabular}
\caption{Variations of (a) $\eta/\rho$ and (b) $\kappa/C_p$ with temperature at 
different values of magnetic field.}\label{smdhcp.1}
\end{center}
\end{figure}

Figure \ref{pl.1} shows the variation of the Prandtl number as a function of temperature for different values of magnetic field and chemical potential. It can be observed that Pr$>$1 and it increases with temperature. The presence of weak magnetic field increases Pr (figure \ref{pl.1}a), whereas the finite chemical potential decreases its magnitude (figure \ref{pl.1}b). The changes of Pr are higher for lower temperatures. The values of the Prandtl number imply that the sound attenuation is mostly governed by the momentum diffusion for the hot QCD matter and it is more pronounced in the presence of weak magnetic field than at finite chemical potential. Here, one can notice that the effect of magnetic field on the Prandtl number is measurable, unlike the effect on $\eta$ and $\zeta$, which can be comprehended as follows. The Prandtl number is the ratio of the momentum diffusion to the thermal diffusion. Both the thermal and momentum diffusions get noticeably affected by the presence of weak magnetic field and the effect on the momentum diffusion (figure \ref{smdhcp.1}a) is found to be larger than the effect on the thermal diffusion (figure \ref{smdhcp.1}b), so their ratio, {\em i.e.} the Prandtl number is noticeably affected by the magnetic field. 

\subsection{Reynolds number}
The viscous behavior of a medium can be understood by studying the Reynolds number, 
\begin{equation}\label{Rl}
{\rm Re}=\frac{Lv}{\eta/\rho}
~,\end{equation}
where ${\eta}/{\rho}$ represents the kinematic viscosity, and $L$ and 
$v$ are the characteristic length and velocity of the flow, 
respectively. Laminar or turbulent nature of the flow is 
specified by the Reynolds number, {\em i.e.} Re requires to be 
much larger than 1 for a turbulent flow while lower values 
correspond to a laminar flow, describing a more viscous fluid \cite{McInnes:NPB921'2017}. The proper time evolution of the thermodynamic quantities in the 
second-order dissipative relativistic fluid dynamics and their dependence on the Reynolds number have been studied in ref. \cite{Muronga:PRC69'2004}. The Reynolds number of quark matter has been estimated to be around 10 using the Kubo formula and NJL model \cite{Fukutome:PTP119'2008}. For initial QGP, (3+1)-dimensional fluid 
dynamical model reports the range of Re to be 3-10 \cite{Csernai:PRC85'2012}, 
whereas its upper bound is estimated to be approximately 20 in the holographic 
model \cite{McInnes:NPB921'2017}. In the present work, the Reynolds number 
for a weakly magnetized hot and dense QCD matter is estimated with $v\simeq 1$ 
and $L=4$ fm. 

\begin{figure}[]
\begin{center}
\begin{tabular}{c c}
\includegraphics[width=6.86cm]{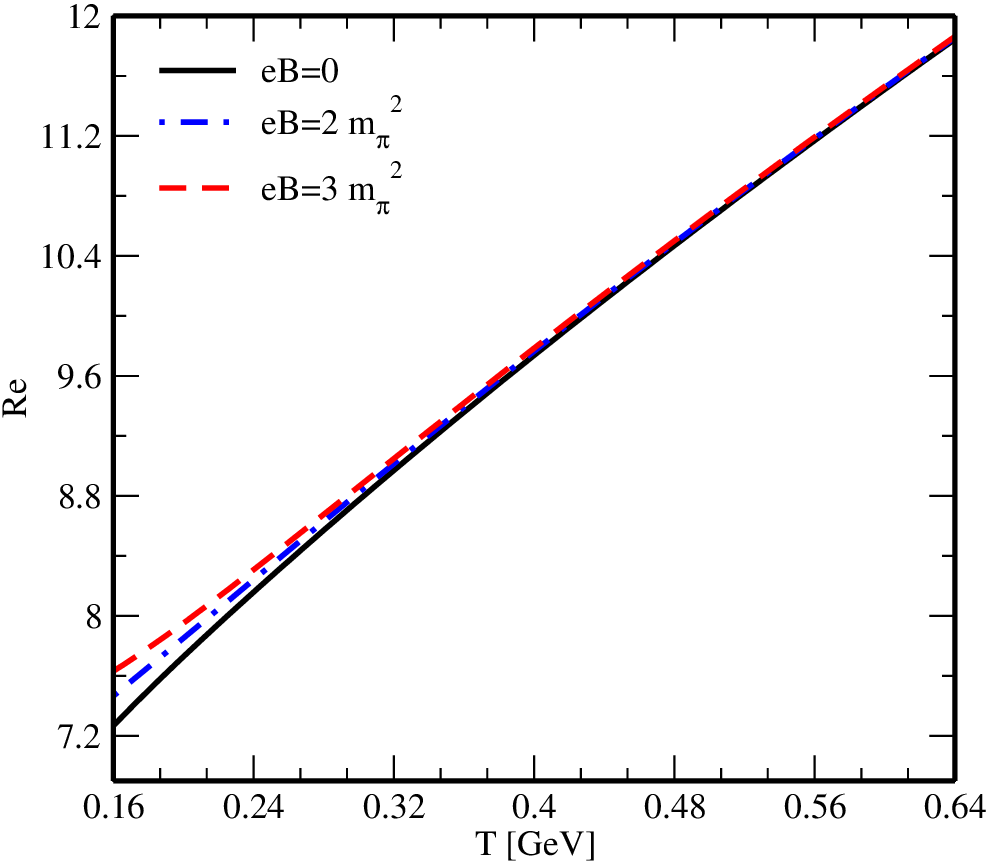}&
\hspace{0.74 cm}
\includegraphics[width=6.86cm]{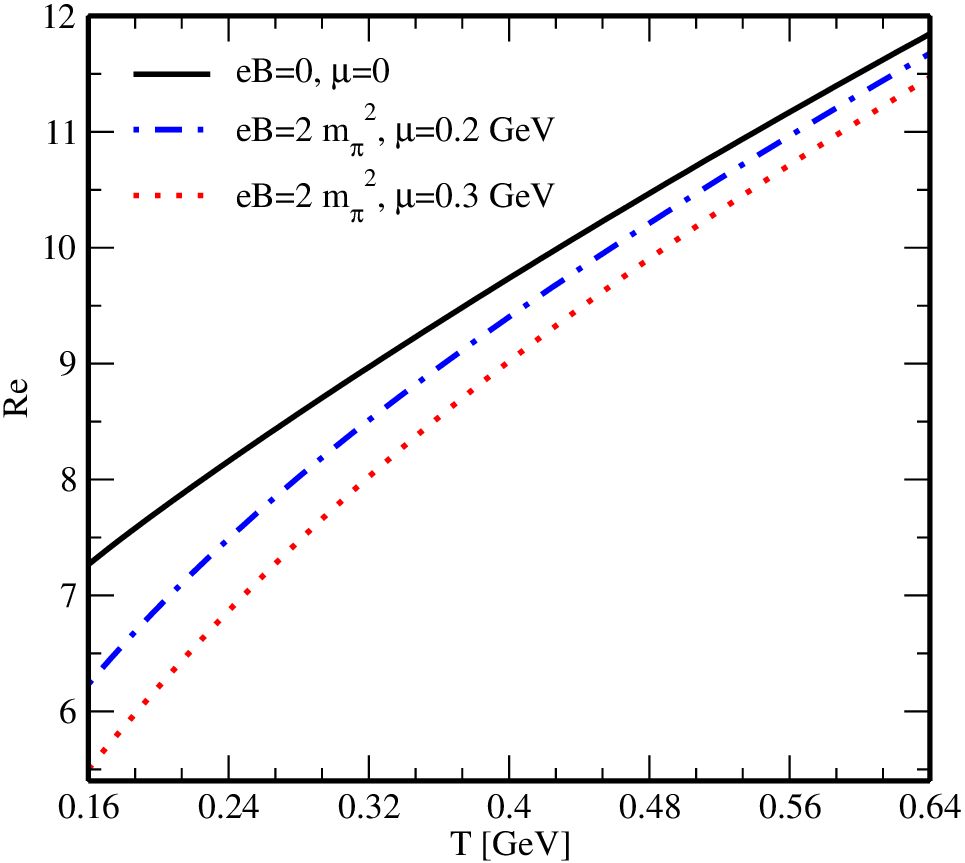} \\
a & b
\end{tabular}
\caption{The variation of the Reynolds number with temperature (a) in the 
presence of weak magnetic field and (b) in the presence of finite 
chemical potential.}\label{rl.1}
\end{center}
\end{figure}

Figure \ref{rl.1} depicts the variation of the Reynolds number with the temperature in the presence of weak magnetic field and finite chemical potential. The Reynolds number is found to increase with the temperature. A small increase in the magnitude of Re is noticed due to the weak magnetic field (figure \ref{rl.1}a), contrary to a large decrease due to the finite chemical potential (figure \ref{rl.1}b). The range of Re is found to be 5.49 - 11.86 in the temperature range, 160 - 640 MeV, indicating that the characteristic length scale of the 
hot QCD system prevails over its kinematic viscosity with the flow remaining laminar. It can be seen that the effect of magnetic field on the Reynolds number is measurable, unlike the effect on $\eta$ and $\zeta$, which can be understood as follows. The Reynolds number is the ratio of the product of characteristic length and velocity of the flow ($Lv$) to the kinematic viscosity. Since $Lv$ has been taken to be constant, the magnitude of effect due to the weak magnetic field is decided by the kinematic viscosity, which is the ratio of the shear viscosity to the mass density. Since the influence of magnetic field on this ratio is noticeable, a measurable effect of magnetic field on the Reynolds number is observed. 

\subsection{Ratios $\eta/s$ and $\zeta/s$}
\begin{figure}[]
\begin{center}
\begin{tabular}{c c}
\includegraphics[width=7cm]{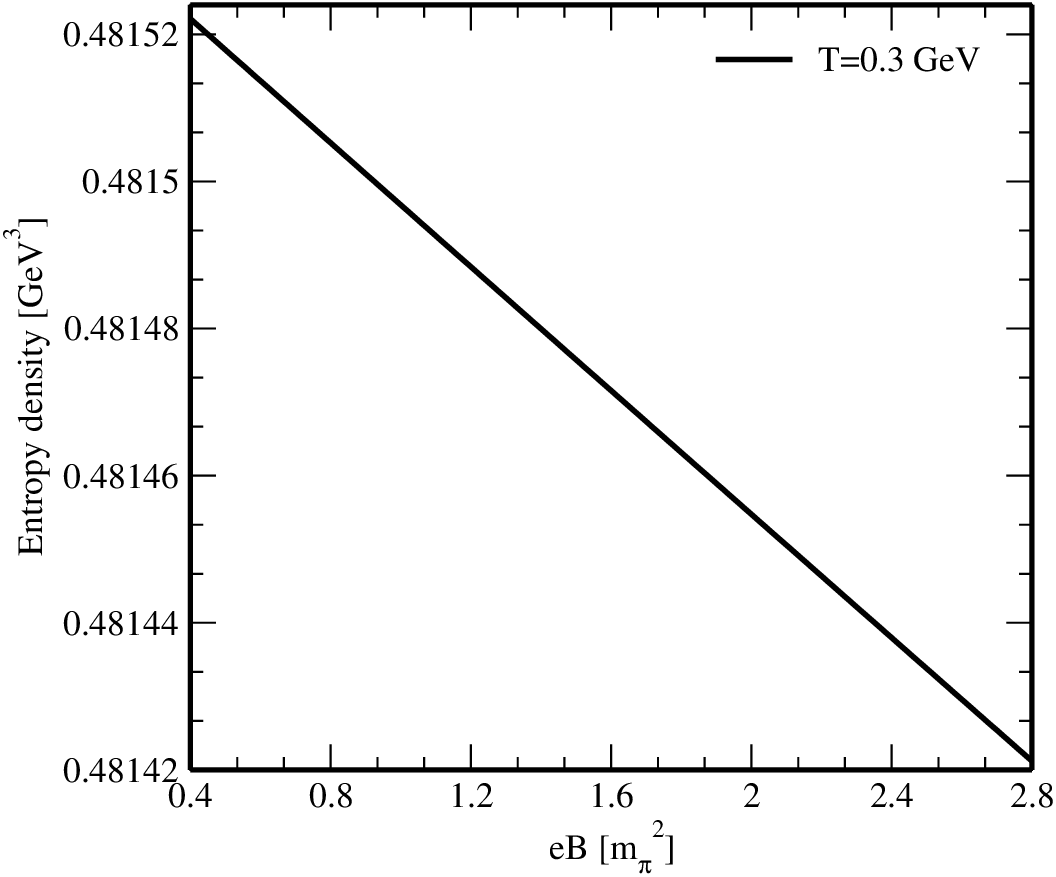}&
\hspace{0.623 cm}
\includegraphics[width=7cm]{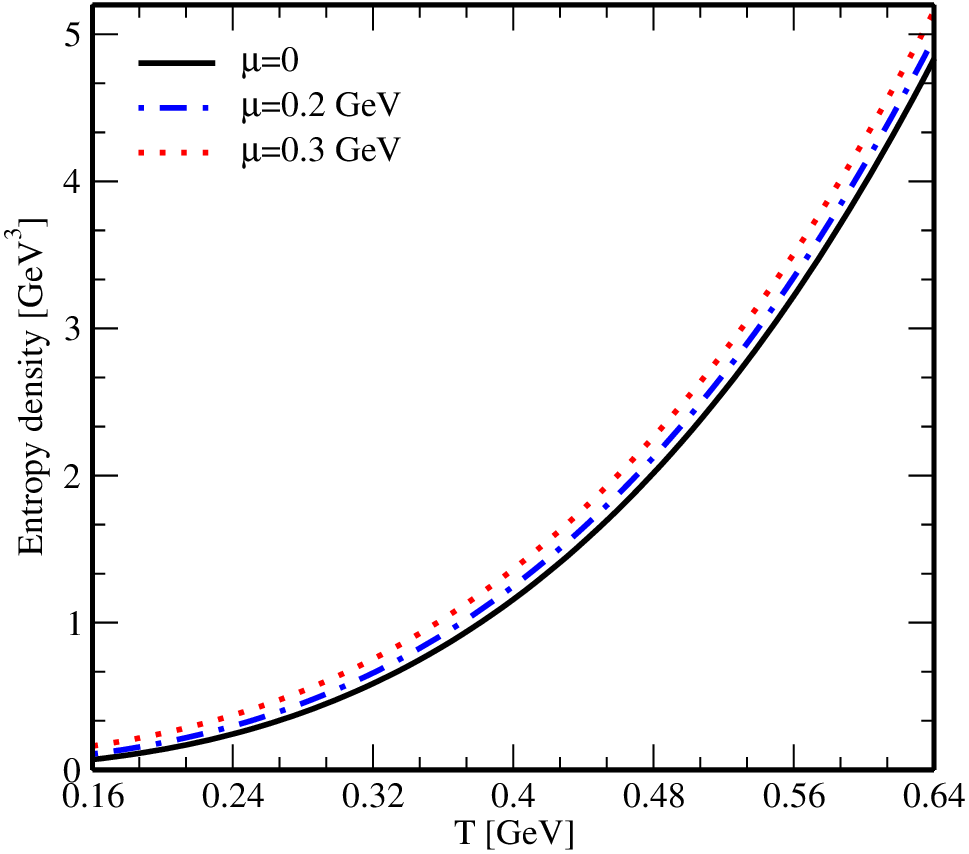} \\
a & b
\end{tabular}
\caption{The variation of the entropy density (a) with magnetic 
field at a fixed temperature and (b) with temperature at 
different values of chemical potential.}\label{ed.1}
\end{center}
\end{figure}

In order to determine the ratios $\eta/s$ and $\zeta/s$, entropy density ($s$) is first evaluated from the energy-momentum tensor and baryon density ($n_B$) using the following equation: 
\begin{eqnarray}\label{E.D.}
S=\frac{u_\mu T^{\mu\nu}u_\nu-\sum_{f}\mu_f n_B-\Delta_{\mu\nu}T^{\mu\nu}/3}{T}
~,\end{eqnarray}
where $n_B$ is defined as
\begin{eqnarray}
n_B &=& \sum_f g_f\int\frac{d^3{\rm p}}{(2\pi)^3}\left(f_f^0-\bar{f}_f^0\right)
.\end{eqnarray}
The entropy density is observed to decrease with an increase of magnetic 
field at a fixed temperature (figure \ref{ed.1}a). On the other hand, an 
increase in the value of entropy density is observed at finite chemical 
potential (figure \ref{ed.1}b). Thus, the presence of magnetic field 
makes the system less disordered, whereas the disorder is larger at 
finite chemical potential. The observations on entropy density, shear and 
bulk viscosities facilitate the exploration of ratios $\eta/s$ and $\zeta/s$. 

\begin{figure}[]
\begin{center}
\begin{tabular}{c c}
\includegraphics[width=7cm]{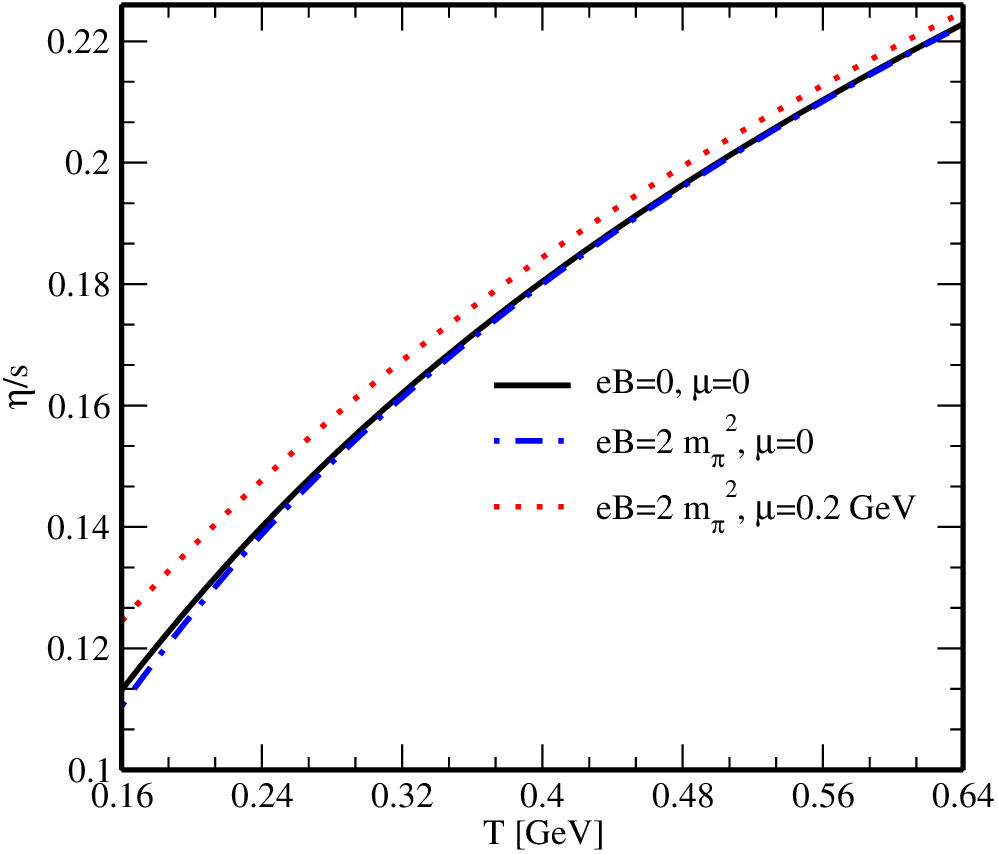}&
\hspace{0.423 cm}
\includegraphics[width=7cm]{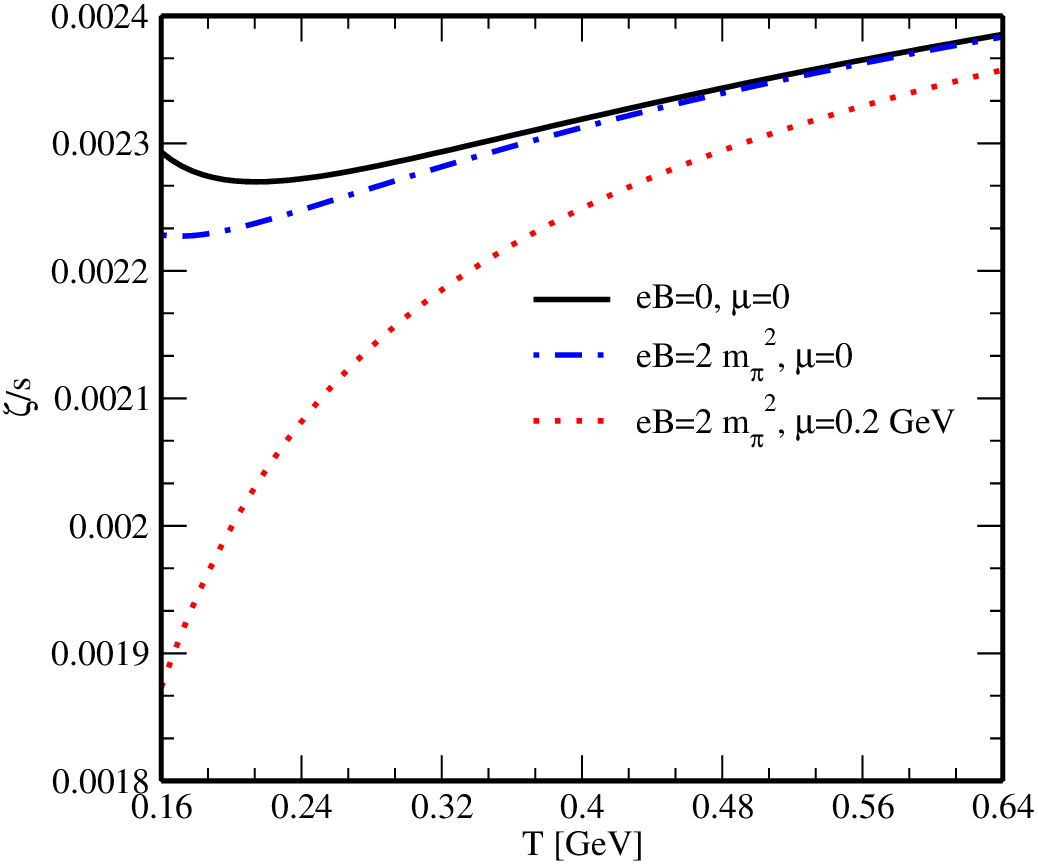} \\
a & b
\end{tabular}
\caption{Variations of (a) ${\eta}/{s}$ and (b) ${\zeta}/{s}$ 
with temperature in the presence of weak magnetic field and 
finite chemical potential.}\label{ratios.1}
\end{center}
\end{figure}

Figures \ref{ratios.1}a and \ref{ratios.1}b display the effects of weak magnetic 
field and finite chemical potential on the variations of $\eta/s$ and 
$\zeta/s$ with temperature, respectively. In a weak magnetic field at 
zero chemical potential, the ratio $\eta/s$ gets slightly decreased and 
becomes nearer to the conjectured lower bound $1/(4\pi)$, specifically at low 
temperatures (figure \ref{ratios.1}a). It can be understood from the 
fact that, both $\eta$ and $s$ get reduced due to the weak magnetic 
field, with the reduction of $\eta$ being more than that of $s$, thus 
resulting in an overall decrease of $\eta/s$ in the said regime. However, 
in the additional presence of chemical potential, $\eta/s$ becomes 
slightly greater than that at $\mu=0$, $eB=0$, but still not 
very far from the lower bound. Thus, the hot QCD matter shows the characteristic 
of a nearly perfect fluid in the said regime. The ratio $\zeta/s$ is found to 
be very small as compared to the ratio $\eta/s$ and it exhibits a nonmonotonic 
behavior at low temperatures (figure \ref{ratios.1}b). Above the phase transition temperature $T_c=0.16$ GeV, there is a broad smooth minimum in the ratio $\zeta/s$ 
and then, this ratio gradually increases at higher temperatures. The presence of 
weak magnetic field slightly decreases the magnitude of $\zeta/s$, which 
corroborates the observations on $\zeta$ and $s$ in the similar environment, 
whereas a comparatively large decrease is observed in the additional presence of 
chemical potential. Unlike $eB=0$, $\mu=0$ case, no nonmonotonic behavior of 
$\zeta/s$ near $T_c$ is found in the presence of both weak magnetic field and 
finite chemical potential. 

The inclusion of magnetic field in the lattice QCD calculations is an emerging area of research. To the best of our knowledge, no lattice QCD results on viscosities are available at finite magnetic field, so it may not be plausible to compare our results on viscosities with the lattice QCD calculations at the equal base. We may however update the lattice QCD results at zero magnetic field. According to the lattice results \cite{Nakamura:PRL94'2005,Karsch:PLB663'2008}, $\eta/s$ becomes minimum and $\zeta/s$ becomes maximum near the phase transition temperature. Compared to the lattice result of $\zeta/s$ in ref. \cite{Karsch:PLB663'2008}, our result in the presence of weak magnetic field is smaller. Lattice calculation for an SU(3) pure gauge model in ref. \cite{Nakamura:PRL94'2005} reports the upper bound for $\eta/s$ of QGP to be 1, and for the temperature range 0.16 GeV - 0.64 GeV, our result on $\eta/s$ in weak magnetic field is slightly less than the lattice result. Another lattice work in ref. \cite{Meyer:PRD76'2007} estimates $\eta/s$ to be nearly 0.102 at $T=1.24T_c$ and 0.134 at $T=1.65T_c$, whereas our weak magnetic field calculation observes slight larger values of $\eta/s$ at these temperatures. Lattice calculation in ref. \cite{Meyer:PRL100'2008} reports a very small value of $\zeta/s$ ($<$0.15) except near $T_c$ and even becomes extremely small away from $T_c$, whereas our weak magnetic field result lies below the lattice result on the ratio $\zeta/s$. The ref. \cite{Astrakhantsev:JHEP1704'2017} has studied the SU(3)-gluodynamics shear viscosity temperature dependence on the lattice and found that for a temperature range $T_c - 1.5T_c$, $\eta/s$ ranges 0.24 - 0.27 approximately. For this temperature range, our calculation estimates $\eta/s$ in the ranges 0.113 - 0.14 at $eB=0$ and 0.11 - 0.139 at $eB\neq 0$. Thus, lattice results lie above our results, and the presence of weak magnetic field shifts $\eta/s$ more towards the lower bound ($1/(4\pi)$), thus making the medium to show nearly perfect fluid characteristics. On the other hand, ref. \cite{Astrakhantsev:PRD98'2018} has studied the SU(3)-gluodynamics bulk viscosity temperature dependence on the lattice and found a very small value of $\zeta/s$ for $T\geq1.1T_c$ and in fact this matches with our result at $eB=0$, whereas the result at weak magnetic field is less than the aforesaid lattice estimation in the same temperature range. Since $\zeta/s$ vanishes for a conformal QGP, the decrease of $\zeta/s$ in weak magnetic field drives the medium towards the conformal symmetry of QGP phase. 

\section{Summary}
The momentum transport properties of a hot and dense QCD matter have been studied 
in terms of the shear and bulk viscosities in the presence of weak magnetic 
field and finite chemical potential using the kinetic theory approach. In general, the emergence of magnetic field breaks the isotropy of the medium and splits the shear viscosity into five components ($\eta_0$, $\eta_1$, $\eta_2$, $\eta_3$ and $\eta_4$) and the bulk viscosity into two components ($\zeta_0$ and $\zeta_1$). Out of these seven components, $\eta_0$ and $\zeta_0$ retain their forms same as in the absence of magnetic field. On the other hand, $\zeta_1$ vanishes, whereas $\eta_1$, $\eta_2$, $\eta_3$ and $\eta_4$ are magnetic field-dependent, out of which $\eta_1$ and $\eta_2$ are dominant shear viscosity coefficients, and $\eta_3$ and $\eta_4$ are called as Hall-type shear viscosity coefficients. In addition, we also calculated the shear ($\eta$) and bulk ($\zeta$) viscosities using the ansatz method in weak magnetic field limit. In this method, the magnetic field-dependence of the bulk viscosity could be seen. This method under the weak magnetic field limit gives only the dominant contribution of the shear viscosity, because we found that $\eta=\eta_2(\approx\eta_1)$. We observed that the presence of weak magnetic field decreases both $\eta$ and $\zeta$, thus reducing the transport of momentum across and along the layer as compared to the zero magnetic field case. The presence of finite chemical potential increases both $\eta$ and $\zeta$. The presence of weak magnetic field makes the Prandtl number (Pr) larger than its value in the absence of both magnetic field and chemical potential, however at finite chemical potential, Pr becomes smaller and in all cases, Pr is  found to be greater than unity. Thus, the sound attenuation is mostly governed by the momentum diffusion and the weak magnetic field makes the dominance of momentum diffusion over thermal diffusion stronger, whereas the chemical potential 
makes this dominance weaker. The Reynolds number (Re) is found to be increased in an 
ambience of weak magnetic field, but it gets decreased at finite chemical potential 
and the flow remains laminar. A meagre decrease due to the weak magnetic field 
and a noticeable increase due to the finite chemical potential in the magnitude 
of $\eta/s$ are observed. The magnitude of $\zeta/s$ and the nonmonotonicity 
in its variation with temperature get waned in the presence of weak magnetic 
field and finite chemical potential. 

\section{Acknowledgment}
One of us (S. R.) is thankful to the Indian Institute of Technology Bombay 
for the Institute postdoctoral fellowship. 

\appendix
\appendixpage
\addappheadtotoc
\begin{appendix}
\renewcommand{\theequation}{A.\arabic{equation}}
\section{Derivation of equation \eqref{deltaf.1}}\label{I.C. of Q.D.F.}
Since magnetic field is taken along $z$-direction, no explicit dependence of magnetic 
field on spatial velocity gradient along $z$-direction can be observed. Now, $J$ 
is calculated as
\be\label{J(1)}
\nonumber J &=& -\beta\tau_f f_f^0\left(1-f_f^0\right)\left[\left\lbrace\omega_f\left(\frac{\partial P}{\partial \varepsilon}\right)-\frac{\rm p^2}{3\omega_f}\right\rbrace\partial_l u^l+p^l\left(\frac{\partial_l P}{\varepsilon+P}-\frac{\partial_l T}{T}\right)\right. \\ && \left.-\frac{Tp^l}{\omega_f}\partial_l\left(\frac{\mu_f}{T}\right)-\frac{p^kp^l}{2\omega_f}W_{kl}\right]
,\ee
where $W_{kl}=\partial_k u_l+\partial_l u_k-\frac{2}{3}\delta_{kl}\partial_j u^j$. After substituting the value of $J$ \eqref{J(1)} in eq. \eqref{eq.2} 
and then simplifying, we obtain 
\be
&& \nonumber -\frac{p_0p_xv_x}{{\rm p}^2}\left\lbrace\omega_f\left(\frac{\partial P}{\partial \varepsilon}\right)-\frac{\rm p^2}{3\omega_f}\right\rbrace\partial_l u^l+\frac{p_kv_xW_{kx}}{2}+\frac{\Gamma_xv_x}{\tau_f}-\omega_c\Gamma_yv_x-qEv_x \\ && \nonumber -\frac{p_0p_yv_y}{{\rm p}^2}\left\lbrace\omega_f\left(\frac{\partial P}{\partial \varepsilon}\right)-\frac{\rm p^2}{3\omega_f}\right\rbrace\partial_l u^l+\frac{p_kv_yW_{ky}}{2}+\frac{\Gamma_yv_y}{\tau_f}+\omega_c\Gamma_xv_y+\omega_c\tau_fqEv_y \\ && +\frac{\Gamma_zv_z}{\tau_f}+\frac{Tp^l}{\omega_f}\partial_l\left(\frac{\mu_f}{T}\right)-p^l\left(\frac{\partial_l P}{\varepsilon+P}-\frac{\partial_l T}{T}\right)=0
,\ee
where $|\mathbf{p}|={\rm p}$ and $\omega_c$ is the cyclotron frequency, $\omega_c=\frac{qB}{\omega_f}$. Equating the coefficients of $v_x$, $v_y$ and $v_z$ on both sides of the above equation, we get 
\be
&&\label{eq.3}-\frac{p_0p_x}{{\rm p}^2}\left\lbrace\omega_f\left(\frac{\partial P}{\partial \varepsilon}\right)-\frac{\rm p^2}{3\omega_f}\right\rbrace\partial_l u^l+\frac{p_kW_{kx}}{2}+\frac{\Gamma_x}{\tau_f}-\omega_c\Gamma_y-qE=0, \\ 
&&\label{eq.4}-\frac{p_0p_y}{{\rm p}^2}\left\lbrace\omega_f\left(\frac{\partial P}{\partial \varepsilon}\right)-\frac{\rm p^2}{3\omega_f}\right\rbrace\partial_l u^l+\frac{p_kW_{ky}}{2}+\frac{\Gamma_y}{\tau_f}+\omega_c\Gamma_x+\omega_c\tau_fqE=0, \\ 
&&\label{eq.5}\frac{\Gamma_z}{\tau_f}=0
.\ee
Now, $\Gamma_x$, $\Gamma_y$ and $\Gamma_z$ can be obtained by solving equations \eqref{eq.3}, \eqref{eq.4} and \eqref{eq.5} as
\be
\label{Gamma(x)}\nonumber\Gamma_x &=& \frac{\tau_f}{1+\omega_c^2\tau_f^2}\frac{p_0p_x}{{\rm p}^2}\left\lbrace\omega_f\left(\frac{\partial P}{\partial \varepsilon}\right)-\frac{\rm p^2}{3\omega_f}\right\rbrace\partial_l u^l+\frac{\omega_c\tau_f^2}{1+\omega_c^2\tau_f^2}\frac{p_0p_y}{{\rm p}^2}\left\lbrace\omega_f\left(\frac{\partial P}{\partial \varepsilon}\right)-\frac{\rm p^2}{3\omega_f}\right\rbrace\partial_l u^l \\ && -\frac{\tau_f}{1+\omega_c^2\tau_f^2}\frac{p_kW_{kx}}{2}-\frac{\omega_c\tau_f^2}{1+\omega_c^2\tau_f^2}\frac{p_kW_{ky}}{2}+\frac{\left(\tau_f-\omega_c^2\tau_f^3\right)qE}{1+\omega_c^2\tau_f^2} ~, \\ 
\label{Gamma(y)}\nonumber\Gamma_y &=& \frac{\tau_f}{1+\omega_c^2\tau_f^2}\frac{p_0p_y}{{\rm p}^2}\left\lbrace\omega_f\left(\frac{\partial P}{\partial \varepsilon}\right)-\frac{\rm p^2}{3\omega_f}\right\rbrace\partial_l u^l-\frac{\omega_c\tau_f^2}{1+\omega_c^2\tau_f^2}\frac{p_0p_x}{{\rm p}^2}\left\lbrace\omega_f\left(\frac{\partial P}{\partial \varepsilon}\right)-\frac{\rm p^2}{3\omega_f}\right\rbrace\partial_l u^l \\ && -\frac{\tau_f}{1+\omega_c^2\tau_f^2}\frac{p_kW_{ky}}{2}+\frac{\omega_c\tau_f^2}{1+\omega_c^2\tau_f^2}\frac{p_kW_{kx}}{2}-\frac{2\omega_c\tau_f^2qE}{1+\omega_c^2\tau_f^2} ~, \\ 
\label{Gamma(z)}\Gamma_z &=& 0
~.\ee
Substituting the values of $\Gamma_x$, $\Gamma_y$ and $\Gamma_z$ in eq. \eqref{ansatz} 
and then simplifying, we get the nonequilibrium part of the quark 
distribution function as follows, 
\be\label{deltaf.(1)}
\nonumber\delta f_f &=& qE\tau_fv_x\beta f_f^0\left(1-f_f^0\right)+v_x\beta f_f^0\left(1-f_f^0\right)\left[\frac{\tau_f}{1+\omega_c^2\tau_f^2}\frac{p_0p_x}{{\rm p}^2}\left\lbrace\omega_f\left(\frac{\partial P}{\partial \varepsilon}\right)-\frac{\rm p^2}{3\omega_f}\right\rbrace\partial_l u^l\right. \\ && \left.\nonumber+\frac{\omega_c\tau_f^2}{1+\omega_c^2\tau_f^2}\frac{p_0p_y}{{\rm p}^2}\left\lbrace\omega_f\left(\frac{\partial P}{\partial \varepsilon}\right)-\frac{\rm p^2}{3\omega_f}\right\rbrace\partial_l u^l-\frac{\tau_f}{1+\omega_c^2\tau_f^2}\frac{p_kW_{kx}}{2}-\frac{\omega_c\tau_f^2}{1+\omega_c^2\tau_f^2}\frac{p_kW_{ky}}{2}\right. \\ && \left.\nonumber+\frac{\left(\tau_f-\omega_c^2\tau_f^3\right)qE}{1+\omega_c^2\tau_f^2}\right]+v_y\beta f_f^0\left(1-f_f^0\right)\left[\frac{\tau_f}{1+\omega_c^2\tau_f^2}\frac{p_0p_y}{{\rm p}^2}\left\lbrace\omega_f\left(\frac{\partial P}{\partial \varepsilon}\right)-\frac{\rm p^2}{3\omega_f}\right\rbrace\partial_l u^l\right. \\ && \left.\nonumber -\frac{\omega_c\tau_f^2}{1+\omega_c^2\tau_f^2}\frac{p_0p_x}{{\rm p}^2}\left\lbrace\omega_f\left(\frac{\partial P}{\partial \varepsilon}\right)-\frac{\rm p^2}{3\omega_f}\right\rbrace\partial_l u^l-\frac{\tau_f}{1+\omega_c^2\tau_f^2}\frac{p_kW_{ky}}{2}\right. \\ && \left.+\frac{\omega_c\tau_f^2}{1+\omega_c^2\tau_f^2}\frac{p_kW_{kx}}{2}-\frac{2\omega_c\tau_f^2qE}{1+\omega_c^2\tau_f^2}\right]
.\ee

\renewcommand{\theequation}{B.\arabic{equation}}
\section{Derivation of equation \eqref{em2.qaqg1}}\label{N.P. of E.M.T.}
The spatial component of eq. \eqref{em1} can be written as
\be\label{em2}
\nonumber\Delta T^{ij} &=& \int\frac{d^3{\rm p}}{(2\pi)^3}p^i p^j \left[\sum_f g_f\frac{\left(\delta f_f+\delta \bar{f}_f\right)}{{\omega_f}}+g_g\frac{\delta f_g}{\omega_g}\right] \\ &=& \Delta T_q^{ij}+\Delta T_{\bar{q}}^{ij}+\Delta T_g^{ij}
.\ee
Using the expression of $\delta f_f$ \eqref{deltaf.1}, the quark part $\Delta T_q^{ij}$ 
in eq. \eqref{em2} is determined as follows, 
\be\label{em2.q}
\nonumber\Delta T_q^{ij} &=& \sum_f g_f\int\frac{d^3{\rm p}}{(2\pi)^3}\frac{p^i p^j}{\omega_f}\delta f_f \\ &=& \nonumber\sum_f g_f\int\frac{d^3{\rm p}}{(2\pi)^3}\frac{p^i p^j}{\omega_f}\beta f_f^0\left(1-f_f^0\right)\left[qE\tau_fv_x+v_x\left(\frac{\tau_f}{1+\omega_c^2\tau_f^2}\frac{p_0p_x}{{\rm p}^2}\left\lbrace\omega_f\left(\frac{\partial P}{\partial \varepsilon}\right)-\frac{\rm p^2}{3\omega_f}\right\rbrace\partial_l u^l\right.\right. \\ && \left.\left.\nonumber+\frac{\omega_c\tau_f^2}{1+\omega_c^2\tau_f^2}\frac{p_0p_y}{{\rm p}^2}\left\lbrace\omega_f\left(\frac{\partial P}{\partial \varepsilon}\right)-\frac{\rm p^2}{3\omega_f}\right\rbrace\partial_l u^l-\frac{\tau_f}{1+\omega_c^2\tau_f^2}\frac{p_kW_{kx}}{2}-\frac{\omega_c\tau_f^2}{1+\omega_c^2\tau_f^2}\frac{p_kW_{ky}}{2}\right.\right. \\ && \left.\left.\nonumber+\frac{\left(\tau_f-\omega_c^2\tau_f^3\right)qE}{1+\omega_c^2\tau_f^2}\right)+v_y\left(\frac{\tau_f}{1+\omega_c^2\tau_f^2}\frac{p_0p_y}{{\rm p}^2}\left\lbrace\omega_f\left(\frac{\partial P}{\partial \varepsilon}\right)-\frac{\rm p^2}{3\omega_f}\right\rbrace\partial_l u^l\right.\right. \\ && \left.\left.\nonumber -\frac{\omega_c\tau_f^2}{1+\omega_c^2\tau_f^2}\frac{p_0p_x}{{\rm p}^2}\left\lbrace\omega_f\left(\frac{\partial P}{\partial \varepsilon}\right)-\frac{\rm p^2}{3\omega_f}\right\rbrace\partial_l u^l-\frac{\tau_f}{1+\omega_c^2\tau_f^2}\frac{p_kW_{ky}}{2}\right.\right. \\ && \left.\left.+\frac{\omega_c\tau_f^2}{1+\omega_c^2\tau_f^2}\frac{p_kW_{kx}}{2}-\frac{2\omega_c\tau_f^2qE}{1+\omega_c^2\tau_f^2}\right)\right]
.\ee
In the weak magnetic field limit, the terms containing $\omega_c$ and 
its higher powers in the numerator can be dropped. Thus, eq. \eqref{em2.q} becomes
\be\label{em2.q1}
\nonumber\Delta T_q^{ij} &=& \nonumber\sum_f g_f\int\frac{d^3{\rm p}}{(2\pi)^3}\frac{p^i p^j}{\omega_f}\beta f_f^0\left(1-f_f^0\right)\left[qE\tau_fv_x+v_x\left(\frac{\tau_f}{1+\omega_c^2\tau_f^2}\frac{p_0p_x}{{\rm p}^2}\left\lbrace\omega_f\left(\frac{\partial P}{\partial \varepsilon}\right)-\frac{\rm p^2}{3\omega_f}\right\rbrace\partial_l u^l\right.\right. \\ && \left.\left.\nonumber -\frac{\tau_f}{1+\omega_c^2\tau_f^2}\frac{p_kW_{kx}}{2}+\frac{\tau_fqE}{1+\omega_c^2\tau_f^2}\right)+v_y\left(\frac{\tau_f}{1+\omega_c^2\tau_f^2}\frac{p_0p_y}{{\rm p}^2}\left\lbrace\omega_f\left(\frac{\partial P}{\partial \varepsilon}\right)-\frac{\rm p^2}{3\omega_f}\right\rbrace\partial_l u^l\right.\right. \\ && \left.\left.\nonumber -\frac{\tau_f}{1+\omega_c^2\tau_f^2}\frac{p_kW_{ky}}{2}\right)\right] \\ &=& \nonumber\sum_f g_f\int\frac{d^3{\rm p}}{(2\pi)^3}\frac{p^i p^j \tau_f}{\omega_f}\beta f_f^0\left(1-f_f^0\right)\left[\frac{2qEv_x}{1+\omega_c^2\tau_f^2}-\frac{1}{1+\omega_c^2\tau_f^2}\frac{p_kp_l}{2\omega_f}W_{kl}\right. \\ && \left.+\frac{1}{1+\omega_c^2\tau_f^2}\left\lbrace\omega_f\left(\frac{\partial P}{\partial \varepsilon}\right)-\frac{\rm p^2}{3\omega_f}\right\rbrace\partial_l u^l\right]
.\ee
Similarly, the antiquark part $\Delta T_{\bar{q}}^{ij}$ is written as
\be\label{em2.aq1}
\nonumber\Delta T_{\bar{q}}^{ij} &=& \nonumber\sum_f g_f\int\frac{d^3{\rm p}}{(2\pi)^3}\frac{p^i p^j \tau_{\bar{f}}}{\omega_f}\beta \bar{f_f}^0\left(1-\bar{f_f}^0\right)\left[\frac{2\bar{q}Ev_x}{1+\omega_c^2\tau_{\bar{f}}^2}-\frac{1}{1+\omega_c^2\tau_{\bar{f}}^2}\frac{p_kp_l}{2\omega_f}W_{kl}\right. \\ && \left.+\frac{1}{1+\omega_c^2\tau_{\bar{f}}^2}\left\lbrace\omega_f\left(\frac{\partial P}{\partial \varepsilon}\right)-\frac{\rm p^2}{3\omega_f}\right\rbrace\partial_l u^l\right]
.\ee
The gluon part $\Delta T_g^{ij}$ retains its form same as that in the absence 
of magnetic field, 
\be\label{em2.g1}
\nonumber\Delta T_g^{ij} &=& g_g\int\frac{d^3{\rm p}}{(2\pi)^3}\frac{p^i p^j\tau_g}{\omega_g}\beta f_g^0\left(1+f_g^0\right)\left[\left\lbrace\omega_g\left(\frac{\partial P}{\partial \varepsilon}\right)-\frac{\rm p^2}{3\omega_g}\right\rbrace\partial_l u^l\right. \\ && \left.-\frac{p^kp^l}{2\omega_g}W_{kl}+p^l\left(\frac{\partial_l P}{\varepsilon+P}-\frac{\partial_l T}{T} \right)\right]
.\ee
Adding equations \eqref{em2.q1}, \eqref{em2.aq1} and \eqref{em2.g1} and then 
simplifying, we get
\be\label{em2.(qaqg1)}
\nonumber\Delta T^{ij} &=& \nonumber\sum_f g_f\int\frac{d^3{\rm p}}{(2\pi)^3}\frac{\beta p^i p^j}{\omega_f}\left[2qEv_x\frac{\tau_f f_f^0\left(1-f_f^0\right)}{1+\omega_c^2\tau_f^2}+2\bar{q}Ev_x\frac{\tau_{\bar{f}}\bar{f_f}^0\left(1-\bar{f_f}^0\right)}{1+\omega_c^2\tau_{\bar{f}}^2}\right. \\ && \left.\nonumber+\left(\frac{\tau_f f_f^0\left(1-f_f^0\right)}{1+\omega_c^2\tau_f^2}+\frac{\tau_{\bar{f}}\bar{f_f}^0\left(1-\bar{f_f}^0\right)}{1+\omega_c^2\tau_{\bar{f}}^2}\right)\left\lbrace\omega_f\left(\frac{\partial P}{\partial \varepsilon}\right)-\frac{\rm p^2}{3\omega_f}\right\rbrace\partial_l u^l\right. \\ && \left.\nonumber -\left(\frac{\tau_f f_f^0\left(1-f_f^0\right)}{1+\omega_c^2\tau_f^2}+\frac{\tau_{\bar{f}}\bar{f_f}^0\left(1-\bar{f_f}^0\right)}{1+\omega_c^2\tau_{\bar{f}}^2}\right)\frac{p_kp_l}{2p_0}W_{kl}\right] \\ && \nonumber +g_g\int\frac{d^3{\rm p}}{(2\pi)^3}\frac{p^i p^j\tau_g}{\omega_g}\beta f_g^0\left(1+f_g^0\right)\left[\left\lbrace\omega_g\left(\frac{\partial P}{\partial \varepsilon}\right)-\frac{\rm p^2}{3\omega_g}\right\rbrace\partial_l u^l\right. \\ && \left.-\frac{p^kp^l}{2\omega_g}W_{kl}+p^l\left(\frac{\partial_l P}{\varepsilon+P}-\frac{\partial_l T}{T} \right)\right]
.\ee

\renewcommand{\theequation}{C.\arabic{equation}}
\section{Derivation of $C_l$, $l=1,2,3,4$}\label{C1C2C3C4}
Making use of the relations $\nabla\cdot\mathbf{V}=0$, $V_{ij}b_ib_j=0$, $b_{ij}v_iv_j=0$, $b_ib_i=1$, $b_{ij}b_i=0$ and $b_{ij}b_j=0$ in eq. \eqref{R.B.T.E.(2)}, we get
\be\label{R.B.T.E.(3)}
\nonumber\beta\omega_fV_{ij}v_iv_jf_0\left(1-f_0\right) &=& -2\omega_c\left[C_1b_{ij}v_jY_{im}^1v_m+C_2b_{ij}v_jY_{im}^2v_m+C_3b_{ij}v_jY_{im}^3v_m
+C_4b_{ij}v_jY_{im}^4v_m\right] \\ && \nonumber+\frac{1}{\tau_f}\left[C_1Y_{ij}^1v_iv_j+C_2Y_{ij}^2v_iv_j+C_3Y_{ij}^3v_iv_j+C_4Y_{ij}^4v_iv_j\right] \\ &=& \nonumber -2\omega_c\left[C_1b_{ij}v_j\left(2V_{im}-2V_{ik}b_kb_m-2V_{mk}b_kb_i\right)v_m\right. \\ && \left.\nonumber+C_2b_{ij}v_j\left(2V_{ik}b_kb_m+2V_{mk}b_kb_i\right)v_m\right. \\ && \left.\nonumber+C_3b_{ij}v_j\left(V_{ik}b_{mk}+V_{mk}b_{ik}-V_{kl}b_{ik}b_mb_l-V_{kl}b_{mk}b_ib_l\right)v_m\right. \\ && \left.\nonumber+C_4b_{ij}v_j\left(2V_{kl}b_{ik}b_mb_l
+2V_{kl}b_{mk}b_ib_l\right)v_m\right] \\ && \nonumber+\frac{1}{\tau_f}\left[C_1\left(2V_{ij}-2V_{ik}b_kb_j-2V_{jk}b_kb_i\right)v_iv_j
\right. \\ && \left.\nonumber+C_2\left(2V_{ik}b_kb_j+2V_{jk}b_kb_i\right)v_iv_j\right. \\ && \left.\nonumber+C_3\left(V_{ik}b_{jk}+V_{jk}b_{ik}-V_{kl}b_{ik}b_jb_l-V_{kl}b_{jk}b_ib_l\right)v_iv_j
\right. \\ && \left.+C_4\left(2V_{kl}b_{ik}b_jb_l+2V_{kl}b_{jk}b_ib_l\right)v_iv_j\right]
.\ee
Now, eq. \eqref{R.B.T.E.(3)} gets further simplified into
\be\label{R.B.T.E.(4)}
\nonumber\beta\omega_fV_{ij}v_iv_jf_0\left(1-f_0\right) &=& -2\omega_c\left[2C_1V_{ik}b_{ij}v_jv_k-2C_1V_{ik}b_{ij}b_kv_j
(\mathbf{b}\cdot\mathbf{v})+2C_2V_{ik}b_{ij}b_kv_j(\mathbf{b}\cdot\mathbf{v})\right. \\ && \left.\nonumber+2C_3V_{ij}v_iv_j-4C_3V_{ij}b_iv_j(\mathbf{b}\cdot\mathbf{v})+2C_4V_{ij}b_iv_j
(\mathbf{b}\cdot\mathbf{v})\right] \\ && \nonumber+\frac{1}{\tau_f}\left[2C_1V_{ij}v_iv_j-4C_1V_{ij}b_iv_j(\mathbf{b}\cdot\mathbf{v})
+4C_2V_{ij}b_iv_j(\mathbf{b}\cdot\mathbf{v})\right. \\ && \left.\nonumber
+C_3V_{ik}b_{jk}v_iv_j+C_3V_{jk}b_{ik}v_iv_j
-2C_3V_{kl}b_{ik}b_lv_i(\mathbf{b}\cdot\mathbf{v})
\right. \\ && \left.+4C_4V_{kl}b_{ik}b_lv_i(\mathbf{b}\cdot\mathbf{v})\right]
.\ee
Comparing the same tensor structures on both sides of eq. \eqref{R.B.T.E.(4)}, we have
\be\label{C1C2C3C4 (1)}
&&-4\omega_cC_3+\frac{2C_1}{\tau_f}=\beta\omega_ff_0\left(1-f_0\right), \\ 
&&-2\omega_c\left(2C_4-4C_3\right)+\frac{1}{\tau_f}\left(-4C_1+4C_2\right)=0, \label{C1C2C3C4 (2)} \\ 
&&-2\omega_c\left(-2C_1+2C_2\right)-\frac{1}{\tau_f}\left(-2C_3+4C_4\right)=0, \label{C1C2C3C4 (3)} \\ 
&&-4\omega_cC_1-\frac{2C_3}{\tau_f}=0 \label{C1C2C3C4 (4)}
.\ee
After solving equations \eqref{C1C2C3C4 (1)}, \eqref{C1C2C3C4 (2)}, \eqref{C1C2C3C4 (3)} and \eqref{C1C2C3C4 (4)}, we get
\be
&&C_1=\frac{\beta\omega_f\tau_ff_0\left(1-f_0\right)}{2\left(1+4\omega_c^2\tau_f^2\right)}, \\ 
&&C_2=\frac{\beta\omega_f\tau_ff_0\left(1-f_0\right)}{2\left(1+\omega_c^2\tau_f^2\right)}, \\ 
&&C_3=-\frac{\beta\omega_f\omega_c\tau_f^2f_0\left(1-f_0\right)}{\left(1+4\omega_c^2\tau_f^2\right)}, \\ 
&&C_4=-\frac{\beta\omega_f\omega_c\tau_f^2f_0\left(1-f_0\right)}{2\left(1+\omega_c^2\tau_f^2\right)}
.\ee

\renewcommand{\theequation}{D.\arabic{equation}}
\section{Thermal conductivity}\label{A.T.C.}
In relativistic hydrodynamics there exist different frames. The freedom to choose a specific frame creates arbitrariness. To avoid arbitrariness, one needs the ``condition of fit'', {\em i.e.} if one chooses the Landau frame, then the condition of fit in the local rest frame requires the ``0'' component of the heat flow four-vector to be zero, {\em i.e.} $Q^0=0$, which can be understood from the fact that in the rest frame of the heat bath or fluid, heat flow four-vector is orthogonal to the fluid four-velocity, {\em i.e.} $Q^\mu u_\mu=0$, where $u_\mu=(1,0,0,0)$ in the local rest frame. Thus in the rest frame of the fluid, the heat flow is purely spatial. This concept has been used in the study of the thermal conductivity. In this way, the results also remain independent of the choice of frame in relativistic hydrodynamics \cite{Gavin:NPA435'1985,Kapusta:PRC85'2012}. 

For a weakly magnetized hot and dense QCD matter, the thermal conductivity is given \cite{Rath:2112.11802} by
\begin{eqnarray}\label{H.C.}
\nonumber\kappa &=& \frac{\beta^2}{6\pi^2}\sum_f g_f\int d{\rm p}~\frac{{\rm p}^4}{\omega_f^2} ~ \left[\frac{\tau_f}{1+\omega_c^2\tau_f^2}\left(\omega_f-h_f\right)^2f_f^0\left(1-f_f^0\right)\right. \\ && \left.+\frac{\tau_{\bar{f}}}{1+\omega_c^2\tau_{\bar{f}}^2}\left(\omega_f-\bar{h}_f\right)^2
\bar{f_f}^0\left(1-\bar{f_f}^0\right)\right]
.\end{eqnarray}

\end{appendix}

\end{document}